\documentclass[twocolumn]{aastex61}

\usepackage{epstopdf}

\usepackage{amsmath}

\newcommand{\Mjup}{\mbox{$M_\mathrm{Jup}$}}
\newcommand{\Msun}{\mbox{$M_{\odot}$}}
\newcommand{\Mearth}{\mbox{$M_{\earth}$}}

\shorttitle{Orbit and Dynamical Mass of Gl 758 B}
\shortauthors{Bowler et al.}

\begin{document}

\title{Orbit and Dynamical Mass of the Late-T Dwarf  Gl 758 B\footnote{Some 
of the data presented herein were obtained at the W.M. Keck Observatory, which is operated as a scientific partnership 
among the California Institute of Technology, the University of California and the National Aeronautics and Space Administration. 
The Observatory was made possible by the generous financial support of the W.M. Keck Foundation.}}

\correspondingauthor{Brendan P. Bowler}
\email{bpbowler@astro.as.utexas.edu}

\author[0000-0003-2649-2288]{Brendan P. Bowler}
\affiliation{McDonald Observatory and the Department of Astronomy, The University of Texas at Austin, Austin, TX 78712, USA}

\author{Trent J. Dupuy}
\affiliation{Gemini Observatory, Northern Operations Center, 670 N. AÕohoku Place, Hilo, HI 96720, USA}
\affiliation{McDonald Observatory and the Department of Astronomy, The University of Texas at Austin, Austin, TX 78712, USA}

\author{Michael Endl}
\affiliation{McDonald Observatory and the Department of Astronomy, The University of Texas at Austin, Austin, TX 78712, USA}

\author{William D. Cochran}
\affiliation{McDonald Observatory and the Department of Astronomy, The University of Texas at Austin, Austin, TX 78712, USA}

\author{Phillip J. MacQueen}
\affiliation{McDonald Observatory and the Department of Astronomy, The University of Texas at Austin, Austin, TX 78712, USA}

\author{Benjamin J. Fulton}
\altaffiliation{Texaco Fellow}
\affiliation{California Institute of Technology, 1200 E. California Blvd., Pasadena, CA 91125, USA}

\author{Erik A. Petigura}
\altaffiliation{Hubble Fellow}
\affiliation{California Institute of Technology, 1200 E. California Blvd., Pasadena, CA 91125, USA}

\author{Andrew W. Howard}
\affiliation{California Institute of Technology, 1200 E. California Blvd., Pasadena, CA 91125, USA}

\author{Lea Hirsch}
\affiliation{Astronomy Department, University of California, Berkeley CA 94720-3411, USA}

\author{Kaitlin M. Kratter}
\affiliation{Department of Astronomy, University of Arizona, Tucson, AZ 85721, USA}

\author{Justin R. Crepp}
\affiliation{Department of Physics, University of Notre Dame, 225 Nieuwland Science Hall, Notre Dame, IN 46556, USA}

\author{Beth A. Biller}
\affiliation{Institute for Astronomy, University of Edinburgh, Blackford Hill View, Edinburgh EH9 3HJ, UK}

\author{Marshall C. Johnson}
\affiliation{Department of Astronomy, The Ohio State University, 140 West 18th Avenue, Columbus, OH 43212 USA}

\author{Robert A. Wittenmyer}
\affiliation{University of Southern Queensland, Computational Engineering and Science Research Centre, Toowoomba, Queensland 4350, Australia}
\affiliation{Australian Centre for Astrobiology, University of New South Wales, Sydney, NSW 2052, Australia}



\begin{abstract}

Gl~758~B is a late-T dwarf orbiting a metal-rich Sun-like star at a projected separation of 
$\rho$~$\approx$~1$\farcs$6 (25~AU).  We present four epochs of astrometry 
of this system with NIRC2 at Keck Observatory spanning 2010 to 2017
together with 630 radial velocities (RVs) of the host star acquired over the past two decades from McDonald Observatory, 
Keck Observatory, and the Automated Planet Finder at Lick Observatory.
The RVs reveal that Gl~758 is accelerating with an evolving rate that varies between 2--5 m s$^{-1}$ yr$^{-1}$,
consistent with the expected influence of the imaged companion Gl~758~B.  
A joint fit of the RVs and astrometry yields a dynamical mass of 42$^{+19}_{-7}$~\Mjup \ for the companion
with a robust lower limit of 30.5~\Mjup \ at the 4-$\sigma$ level.
Gl~758~B is on an eccentric orbit ($e$~=~0.26--0.67 at 95\% confidence) with a semimajor axis of 
$a$~=~$21.1_{-1.3}^{+2.7}$~AU and an orbital period of $P$~=~$96_{-9}^{+21}$ yr,
which takes it within $\approx$9~AU from its host star at periastron passage.
Substellar evolutionary models generally underpredict the mass of Gl~758~B
for nominal ages of 1--6~Gyr that have previously been adopted for the host star.  
This discrepancy can be reconciled if the system is older---which 
is consistent with activity indicators and recent isochrone fitting of the host star---or 
alternatively if the models are systematically overluminous by $\approx$0.1--0.2~dex.
Gl~758~B is currently the lowest-mass directly imaged companion inducing a measured acceleration on its host star.
In the future, bridging RVs and high-contrast imaging 
with the next generation of extremely large telescopes and space-based facilities
will open the door to the first dynamical mass measurements of imaged exoplanets.

\end{abstract}

\keywords{brown dwarfs --- stars: individual (Gl 758)}




\section{Introduction}{\label{sec:intro}}

Brown dwarfs and giant planets radiatively cool over time and follow mass-luminosity-age relationships.
Two quantities---usually luminosity and age---are needed to infer the third using substellar evolutionary models, 
which is how the masses of all directly imaged
exoplanets and the vast majority of brown dwarfs are estimated.  
Fundamental tests of these cooling models require measurements of all three parameters, making
model-independent dynamical masses especially valuable for the subset of brown dwarfs with well-constrained ages.
These rare benchmarks with measured luminosities, ages, and dynamical masses
have shown that
widely used hot-start evolutionary models systematically over-predict masses by up to 25\%,
a discrepancy that may originate from diverse accretion histories or 
incomplete modeling of cloud evolution from L to T spectral classes
(\citealt{Dupuy:2009jq}; \citealt{Crepp:2012eg}; \citealt{Dupuy:2014iz}), although the exact source 
of this deviation remains unclear.
Expanding these tests to even lower brown dwarf masses---and ultimately into the planetary regime---will 
enable precision tests of giant planet evolutionary, atmospheric, and formation models.

An especially useful class of benchmark brown dwarfs are those orbiting at wide enough separations to be identified and characterized
with direct imaging, but close enough that their influence on their host stars can be readily measured with radial velocities (RVs).
One of the legacy products of long-baseline precision RV planet searches operating over the past twenty years 
is the identification of systems exhibiting shallow accelerations.
These ``dynamical beacons'' point to distant stellar, substellar, or planetary companions and are excellent targets for follow-up high-contrast imaging to determine the nature of the perturbing body (e.g., \citealt{Bowler:2016jk}).

Four high-mass brown dwarfs have been recovered with high-contrast imaging 
based on long-term RV trends from their host stars:
HR 7672 B (\citealt{Liu:2002fx}), 
HD 19467 B (\citealt{Crepp:2014ce}),
HD 4747 B (\citealt{Crepp:2016fg}), and
HD 4113 C (\citealt{Cheetham:2017wb}).
These benchmark brown dwarfs have mid-L to late-T spectral types, dynamical masses between 50--70 \Mjup,
and ages between 1--7 Gyr.
HD 4113 C is especially peculiar; the inferred mass of this late-T dwarf companion is about a factor of two lower
than its dynamical mass, suggesting that it may be an unresolved binary T dwarf.
In addition, over two dozen brown dwarfs in close binaries have had their masses measured through patient orbit monitoring campaigns
(e.g., \citealt{Liu:2008ib}; \citealt{Konopacky:2010kr}; \citealt{Dupuy:2017ke}).  However, their ages are usually difficult 
to independently constrain, even with component-resolved spectroscopy, unless these binaries also happen to be gravitationally bound
to a host star (\citealt{Mccaughrean:2004ey}; \citealt{Potter:2002ie}; \citealt{Ireland:2008kr}; \citealt{Dupuy:2009jq}).

In this work, we present a dynamical mass measurement of the late-T dwarf Gl 758 B based on new 
high-contrast imaging observations from Keck/NIRC2 together with 630 RVs of the host star
from McDonald Observatory, Keck Observatory, and the Automated Planet Finder 
gathered over the past 20 years.
Orbital motion is evident in all datasets; Gl 758 B displays changes
in position angle (P.A.) and separation in our imaging data, and our
precision RVs show clear signs of a shallow acceleration with slight curvature.
With a dynamical mass of 42$^{+19}_{-7}$~\Mjup, Gl 758 B is likely to be the lowest-mass imaged companion inducing a 
measured acceleration on its host star and serves as a valuable test for substellar evolutionary models.

Section~\ref{sec:overview} provides an overview of the Gl 758 system and summarizes the physical properties 
of the late-T dwarf companion.  Section~\ref{sec:rvobs} describes the precision RV observations of the host star.
Our Keck observations, PSF subtraction, and astrometric measurements are reported in Section~\ref{sec:astobs}.
The joint Keplerian fit to the RV and astrometric data can be found in Section~\ref{sec:orbit}.
Finally, we compare the results to predictions from evolutionary models in Section~\ref{sec:discussion}
and conclude in Section~\ref{sec:conclusions}.

\section{Overview of the Gl 758 System}{\label{sec:overview}}

Gl 758 (=HD182488, HR 7368) is a bright ($V$=6.3~mag) G8 star located at 15.66 $\pm$ 0.09 pc (\citealt{GaiaCollaboration:2016gd}).
Activity, lithium, and kinematics of this star all point to an age of 3$^{+3}_{-2}$~Gyr, implying a mass
of about 0.97~\Msun, and enhanced metallicity of [Fe/H] $\approx$ +0.2 dex 
(see \citealt{Vigan:2016gq} for a thorough summary).

The brown dwarf companion Gl 758 B was first discovered by \citet{Thalmann:2009ca} 
as part of the SEEDS high-contrast imaging survey (\citealt{Tamura:2016jg}).
Further photometric and spectroscopic characterization by \citet{Currie:2010ju}, 
\citet{Janson:2011dh}, \citet{Vigan:2016gq}, and \citet{Nilsson:2017hm}
established it as a late-T dwarf (T7--T8) with a model-based mass between 10--40 \Mjup \ 
and an effective temperature of 600--750 K.
\citet{Vigan:2016gq} find that no empirical or model template accurately reproduces
the ensemble of photometry for this object, possibly due to an enhanced metallicity.
Although Gl 758 B has only completed a small fraction of its orbit, the most likely orbital solutions
relying solely on astrometry 
suggest it is eccentric with a semimajor axis between about 20--60 AU.
\citet{Nilsson:2017hm} propose that the acceleration induced by Gl 758 B should be
measurable on the host star, but no evidence of a trend was observed by 
\citet{Vigan:2016gq} using RVs from the ELODIE spectrograph and Lick Observatory,
most likely due to the relatively large uncertainties of these datasets.

\section{Radial Velocity Observations}{\label{sec:rvobs}}

\subsection{Harlan J. Smith Telescope/Tull Spectrograph}{\label{sec:mcdonald}}

\begin{figure}
  \vskip -.5 in
  \hskip -.8 in
  \resizebox{5in}{!}{\includegraphics{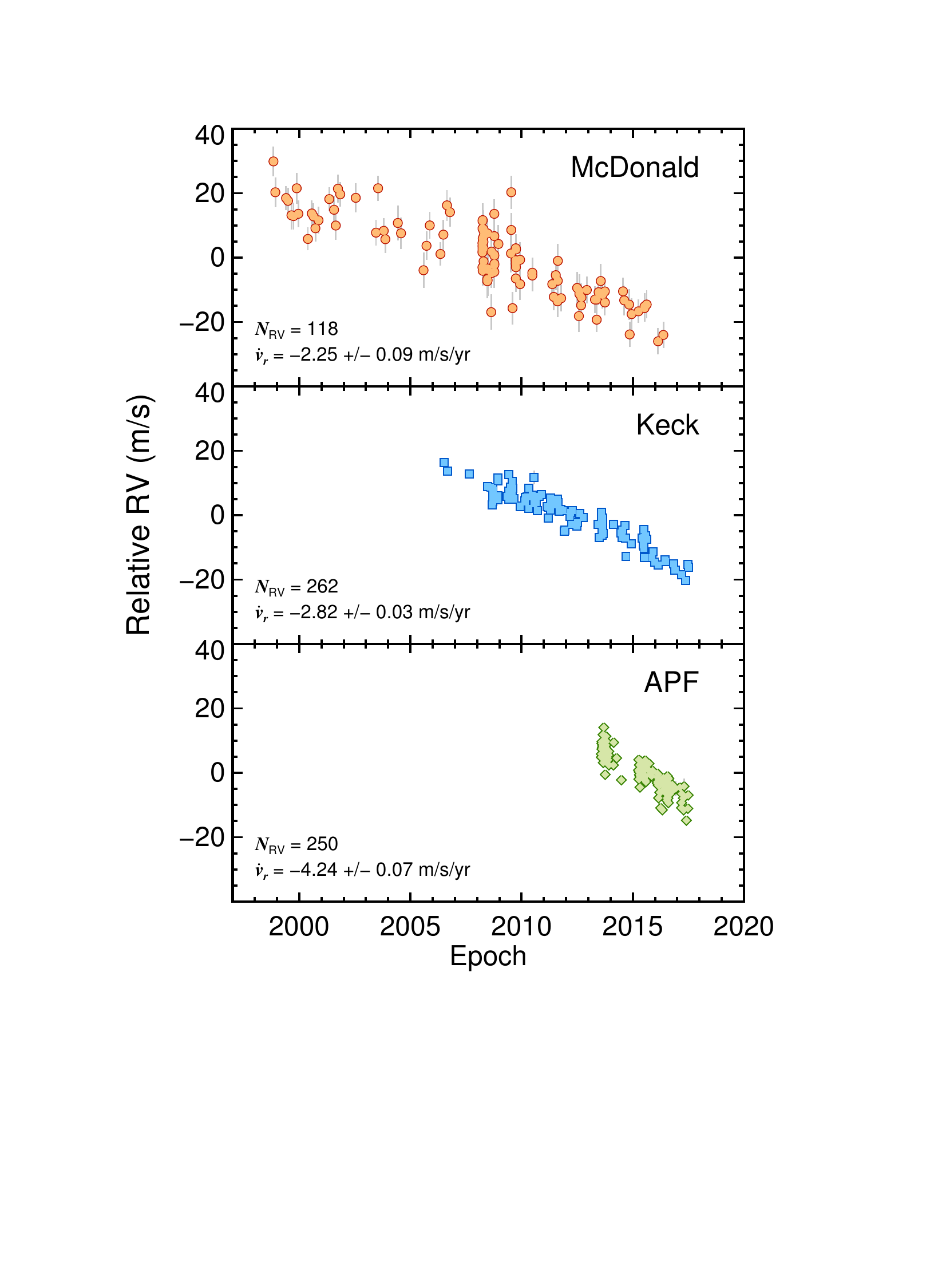}} 
  \vskip -1.7 in
  \caption{Relative RVs of Gl 758 from the Tull Spectrograph at McDonald Observatory's 
  Harlan J. Smith Telescope (top), HIRES at Keck Observatory (middle), and the Levy Spectrograph
  at APF (bottom).  A linear fit to each dataset shows an evolving and 
  steepening slope at later epochs--- listed in the bottom-left corner of each panel--- 
  implying recent changes in the radial acceleration of Gl 758. \label{fig:rvs} } 
\end{figure}

Gl 758 was included in the target sample of the long-duration RV planet search at 
McDonald Observatory (e.g., \citealt{Cochran:1997ta}, \citealt{Endl:2016kk}). 
The Tull Coud\'e spectrograph (\citealt{Tull:1995tn}) was used at the Harlan J. Smith 2.7\,m 
telescope in combination with an I$_2$ cell in the light path 
to obtain precise differential RVs. We commenced observations of Gl~758 on 
1998 November 4th, and have accumulated 118 precise RV measurements 
over the past 19 years. 
Beginning in 2009 an exposure meter was used to provide the optimal exposure level 
and compute precise barycentric corrections.
We measure precise RVs from these spectra using 
our {\it Austral} I$_2$ cell data code (\citealt{Endl:2000ui}).  
The exposure times of the GJ~758 spectra range from 365 seconds to 1200 seconds, primarily controlled by atmospheric seeing conditions to reach 
the desired SNR of $\sim$300 per pixel in the I$_2$ bandpass (500 to 650 nm). 
The RV data have a total rms of 13\,m\,s$^{-1}$ and a median uncertainty of 4.6\,m\,s$^{-1}$. 
Our measurements are listed in Table~\ref{tab:rvs} and displayed in Figure~\ref{fig:rvs}.

\startlongtable
\begin{deluxetable}{lccc}
\renewcommand\arraystretch{0.9}
\tabletypesize{\small}
\setlength{ \tabcolsep } {.1cm} 
\tablewidth{0pt}
\tablecolumns{4}
\tablecaption{Relative Radial Velocities \label{tab:rvs}}
\tablehead{
       \colhead{Date} &  \colhead{RV}  & \colhead{$\sigma_\mathrm{RV}$} & \colhead{Obs.}   \\
   \colhead{(BJD$_\mathrm{TDB}$)}    &  \colhead{(m s$^{-1}$)}       & \colhead{(m s$^{-1}$)}     &      
        }   
\startdata
\cutinhead{McDonald Observatory}
       2451121.66132    &   29.88    &    4.63    &    McD   \\
       2451152.55410    &   20.29    &    4.57    &    McD   \\
       2451328.93255    &   18.48    &    3.78    &    McD   \\
       2451360.93134    &   17.64    &    4.02    &    McD   \\
       2451417.85387    &   13.14    &    4.49    &    McD   \\
       2451452.68967    &   12.89    &    4.15    &    McD   \\
       2451503.57734    &   21.52    &    4.73    &    McD   \\
       2451530.54249    &   13.60    &    4.10    &    McD   \\
       2451686.95063    &    5.80    &    3.57    &    McD   \\
       2451751.77696    &   13.70    &    3.98    &    McD   \\
\multicolumn{4}{c}{$\cdots$} \\
\cutinhead{Keck Observatory}
       2453927.88034    &   19.41    &    0.99    &    Keck   \\
       2453927.88136    &   18.40    &    1.03    &    Keck   \\
       2453927.88237    &   17.27    &    1.13    &    Keck   \\
       2453982.88766    &   17.25    &    0.90    &    Keck   \\
       2453982.88864    &   14.97    &    0.96    &    Keck   \\
       2453982.88965    &   15.15    &    0.95    &    Keck   \\
       2454338.96476    &   14.86    &    1.14    &    Keck   \\
       2454642.00375    &   11.02    &    1.22    &    Keck   \\
       2454689.89631    &   10.76    &    1.15    &    Keck   \\
       2454717.84714    &    5.36    &    1.22    &    Keck   \\
\multicolumn{4}{c}{$\cdots$} \\
\cutinhead{APF}
       2456505.87953    &    6.17    &    0.89    &    APF   \\
       2456506.79254    &    5.16    &    0.93    &    APF   \\
       2456515.81319    &    7.68    &    1.00    &    APF   \\
       2456516.85698    &    9.41    &    0.87    &    APF   \\
       2456517.75297    &   10.05    &    0.93    &    APF   \\
       2456518.77451    &    9.55    &    0.89    &    APF   \\
       2456525.74145    &    8.28    &    1.04    &    APF   \\
       2456526.74485    &    9.92    &    0.87    &    APF   \\
       2456530.79148    &    8.83    &    1.05    &    APF   \\
       2456534.75735    &   12.16    &    0.92    &    APF   \\
\multicolumn{4}{c}{$\cdots$} \\
\enddata
\tablecomments{Table 2 is published in its entirety in the machine-readable format.
      A portion is shown here for guidance regarding its form and content.}
\end{deluxetable}


\subsection{Keck/HIRES}{\label{sec:hires}}

Gl 758 has been continuously monitored with
the High Resolution Echelle Spectrometer (HIRES; \citealt{Vogt:1994tb}) at the Keck I telescope since 2006.
This star was initially targeted as an RV standard for FGK stars in the $Kepler$ field due to its 
proximity on the sky to the $Kepler$ footprint and its stable RVs, but after several years of coverage it 
became apparent that Gl 758 was undergoing a shallow radial acceleration  (G. Marcy, private communication).
Altogether 262 spectra were gathered following the standard setup, observing strategy, 
and procedure for measuring relative RVs implemented by the 
California Planet Search program (\citealt{Howard:2010dia}).
An iodine cell is mounted in the optical path before the slit entrance
to provide a set of stable reference lines (\citealt{Marcy:1992ix}),
an exposure meter is used to consistently achieve a SNR of about 225 per reduced pixel near 550 nm,
and relative RVs are extracted by forward modeling the stellar and iodine spectra 
convolved with the instrument line spread function (\citealt{Valenti:1995bk}).
The median uncertainty from these measurements is 1.2 m s$^{-1}$.
A secular trend of --2.82 $\pm$ 0.03 m s$^{-1}$ yr$^{-1}$ is apparent 
in our Keck RVs (Figure~\ref{fig:rvs}).  This slope is slightly steeper
than the trend from McDonald Observatory, suggesting a recent change
in the acceleration.
This is readily apparent by considering only the latest HIRES data, which 
exhibits a slope of --3.08 m s$^{-1}$ yr$^{-1}$ since 2010, and --4.15 m s$^{-1}$ yr$^{-1}$ since 2013.5.
Our HIRES RVs and uncertainties are listed in Table~\ref{tab:rvs}.

\subsection{Automated Planet Finder Telescope/Levy Spectrograph}{\label{sec:apf}}

Observations of Gl 758 have been carried out autonomously at the 2.4-m Automated Planet Finder (APF) telescope 
at Lick Observatory since 2013.  250 echelle spectra were gathered as part of APF's automated Doppler search for
rocky planets (\citealt{Fulton:2015gj})  with the Levy Spectrograph, which employs
an iodine cell to measure precise relative RVs (\citealt{Vogt:2014wxa}).  Each spectrum spans 3740--9700~\AA \ at a resolving power of $\approx$100,000 with the 1$''$ decker.
RVs are measured by forward-modeling 848 spectral regions, and the resulting variance
is adopted as the RV uncertainty.  The median instrumental precision of these measurements 
for Gl 758 is 1.4 m s$^{-1}$.
The APF RVs reveal an acceleration that is significantly steeper than the 
McDonald and Keck RVs (Figure~\ref{fig:rvs}).  The linear trend
from APF is --4.24 $\pm$ 0.07 m s$^{-1}$ yr$^{-1}$, 
indicating a substantial evolution in recent years.
Fortunately, these evolving slopes provide curvature that can better constrain the
orbit and dynamical mass of the companion.
Our APF RVs and uncertainties are listed in Table~\ref{tab:rvs}.

\section{Astrometric Observations}{\label{sec:astobs}}

\subsection{Keck/NIRC2 Adaptive Optics Imaging}{\label{sec:aoimaging}}

We observed Gl 758 with the NIRC2 camera in its narrow mode (10$\farcs$2$\times$10$\farcs$2 field of view)
using natural guide star adaptive optics 
(\citealt{Wizinowich:2013dz}) at the Keck II telescope on four occasions: UT 2010 May 02, UT 2013 July 03, UT 2016 June 27, and UT 2017 October 10.
All observations were taken in the pupil-tracking mode to employ the angular differential imaging 
(ADI) method (\citealt{Marois:2006df}).
The star was placed behind the partly-transparent 600~mas coronagraph, which has an 
attenuation of 7.51 $\pm$ 0.14 mag at 1.6~$\mu$m (\citealt{Bowler:2015ja}) and enables
precise image registration.
The total on-source integration time of our observations was 49 min, 14 min, 35 min, and 30 min 
for our 2010, 2013, 2016, and 2017 epochs, respectively, and the total sky rotation of these sequences was
65$^{\circ}$, 13$^{\circ}$, 59$^{\circ}$, and 48$^{\circ}$.
Images were corrected for cosmic rays and bad pixels, then dark subtracted and flat fielded.
Details about the observations are summarized in Table~\ref{tab:obs}.  Further processing of the
images is described in Section~\ref{sec:astrometry}.

\begin{deluxetable*}{lccccccc}
\renewcommand\arraystretch{0.9}
\tabletypesize{\small}
\setlength{ \tabcolsep } {.1cm} 
\tablewidth{0pt}
\tablecolumns{8}
\tablecaption{Keck/NIRC2 Adaptive Optics Imaging of Gl 758 \label{tab:obs}}
\tablehead{
       \colhead{UT Date} &  \colhead{Epoch}  & \colhead{$N$$\times$Coadds$\times$$t_\mathrm{exp}$}  & \colhead{Filter\tablenotemark{a}}  & \colhead{Sep.}  & \colhead{P.A.}  &  \colhead{Detection}  &  \colhead{PSF}  \\
      &  \colhead{(UT)}       & \colhead{(s)}                  & \colhead{}                      &  \colhead{($''$)}       & \colhead{($^{\circ}$)} &  \colhead{SNR}  & \colhead{Ref.}           
        }   
\startdata
\cutinhead{Gl 758 B}
2010 May 02  & 2010.333  &  98 $\times$ 6 $\times$ 5  &  $CH_4s$+cor600  &  1.8480 $\pm$ 0.0018  &  200.6 $\pm$ 0.3  & 24.5 & 1  \\
2013 Jul 03  & 2013.502  &  28 $\times$ 30 $\times$ 1  &  $K_S$+cor600  & 1.743  $\pm$ 0.002  &  205.7 $\pm$  0.2   & 10.5  & 2 \\
2016 Jun 27  & 2016.489  &  70 $\times$ 6 $\times$ 5  &  $H$+cor600  &  1.6256 $\pm$ 0.0019  &  210.3 $\pm$ 0.4 & 20.9  &  1  \\
2017 Oct 10  & 2017.773  &  60 $\times$ 6 $\times$ 5  &  $H$+cor600  &  1.588 $\pm$ 0.002  &  213.5 $\pm$ 0.3   &  38.5  & 3  \\
\cutinhead{bkg1}
2010 May 02  & 2010.333  &  98 $\times$ 6 $\times$ 5  &  $CH_4s$+cor600  &  1.390 $\pm$ 0.002  &  222.6 $\pm$ 0.3  & 8.4  & 1 \\
2013 Jul 03  & 2013.502  &  28 $\times$ 30 $\times$ 1  &  $K_S$+cor600  &  1.931 $\pm$ 0.002  &  216.1 $\pm$ 0.2   & 11.5  &  2 \\
2016 Jun 27  & 2016.489  &  70 $\times$ 6 $\times$ 5  &  $H$+cor600  &  2.485 $\pm$ 0.002  &  213.5 $\pm$ 0.4   & 51.1  &  1 \\
2017 Oct 10  & 2017.773  &  60 $\times$ 6 $\times$ 5  &  $H$+cor600  &  2.642 $\pm$ 0.002  &  212.7 $\pm$ 0.3   &  49.1  & 3  \\
\cutinhead{bkg2}
2016 Jun 27  & 2016.489  &  70 $\times$ 6 $\times$ 5  &  $H$+cor600  &  1.4585 $\pm$ 0.0017  &  177.7 $\pm$ 0.4  & 10.6  &  1  \\
2017 Oct 10  & 2017.773  &  60 $\times$ 6 $\times$ 5  &  $H$+cor600  &  1.6246 $\pm$ 0.0018  &  179.9 $\pm$ 0.3   &  31.5  &  3  \\
\enddata
\tablenotetext{a}{``cor600'' refers to the 600 mas-diameter focal plane coronagraph.}
\tablecomments{NIRC2 astrometric PSF reference stars make use of the following datasets: 
(1) $H$ band imaging of PYC11519+0731 from 2012 May 22 UT (\citealt{Bowler:2015ch}); 
(2) $K_S$ band imaging of 2M22362452+4751425 from 2015 August 27 UT (\citealt{Bowler:2017hq});
(3) $H$ band imaging of HD 109461 from 2017 October 10 UT.}
\end{deluxetable*}

\subsection{PSF Subtraction and Astrometry}{\label{sec:astrometry}}

\begin{figure*}
  \vskip -1.2 in
  \hskip -0.5 in
  \resizebox{9.5in}{!}{\includegraphics{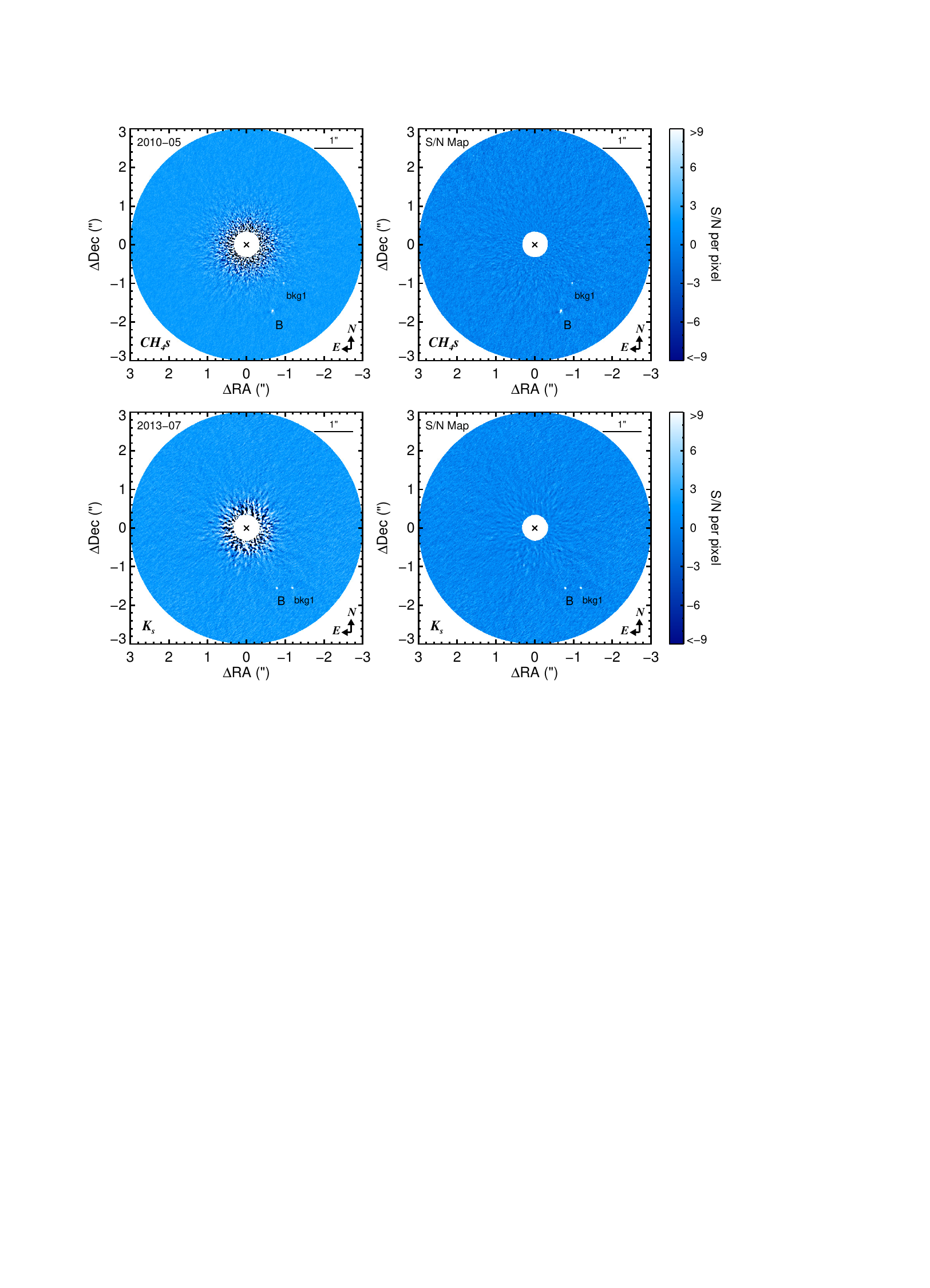}}
  \vskip -6.2 in
  \caption{Final PSF-subtracted images of Gl 758 taken with Keck/NIRC2 in 2010 and 2013 (upper and lower panels, respectively).  Images on the left are the processed frames in units of flux (DN s$^{-1}$), and images on the right are the corresponding signal-to-noise maps.  The color bar on the far right corresponds to intensities in the SNR map.  All images are oriented so that North is up and East is to the left.  Most of the proper motion of Gl 758 is in the positive declination direction, so the background source ``bkg1'' move downward over time relative to Gl 758. \label{fig:nirc2a} } 
\end{figure*}

\begin{figure*}
  \vskip -7 in
  \hskip -0.5 in
  \resizebox{9.5in}{!}{\includegraphics{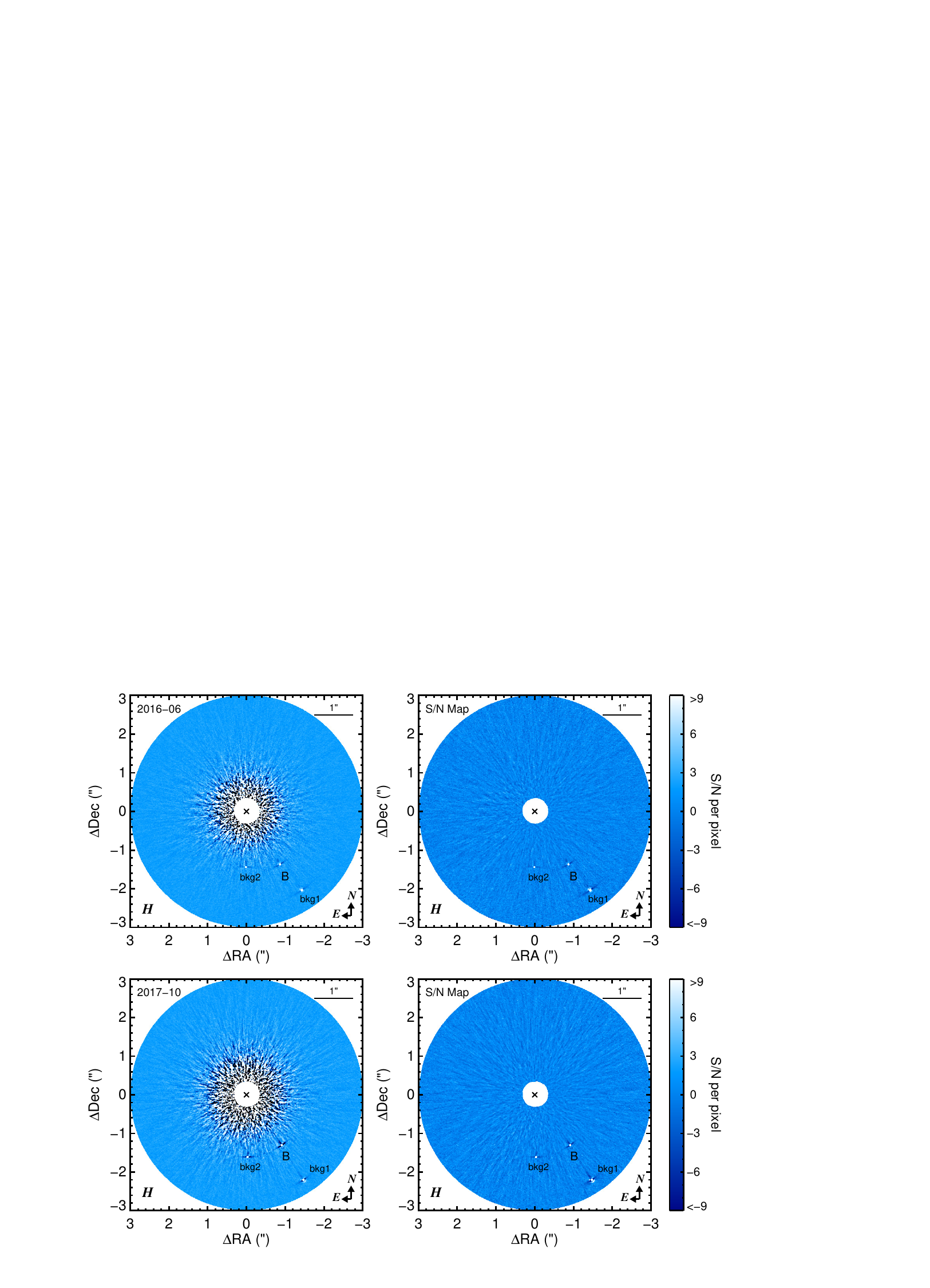}}
  \vskip -.3 in
  \caption{PSF-subtracted images of Gl 758 taken with Keck/NIRC2 in 2016 and 2017 (upper and lower panels, respectively; see Figure~\ref{fig:nirc2a} for details).  All images are oriented so that North is up and East is to the left.  The background sources ``bkg1'' and ``bkg2'' are marked and move downward over time with respect to Gl 758, while the bound companion Gl 758 B orbits in a counter-clockwise direction. \label{fig:nirc2b} } 
\end{figure*}

PSF subtraction for the NIRC2 imaging data is carried out with the Locally-Optimized Combination of Images algorithm (LOCI; \citealt{Lafreniere:2007bg}) 
using the ADI processing pipeline described in \citet{Bowler:2015ja}.  
Images were individually corrected for geometric distortions by bilinearly interpolating pixel values to the 
rectified locations based on the solution from 
\citet{Yelda:2010ig} for observations taken prior to April 2015 when the Keck II AO system was re-aligned, and 
the solution from \citet{Service:2016gk} was used for observations taken after pupil realignment.
Each frame was then registered by fitting a 2D elliptical Gaussian to the 
host star located behind the partly-transparent coronagraph spot.  
Two reductions were carried out using aggressive and conservative implementations 
of LOCI by varying the angular tolerance parameter used to select
PSF templates ($N_{\delta}$).  
Two point sources are recovered with high significance in the 2010 and 2013 epochs (Gl 758 B
and bkg1; Figure~\ref{fig:nirc2a}), and 
three point sources are recovered in the 2016 and 2017 epochs (Gl 758 B, bkg1, and bkg2; Figure~\ref{fig:nirc2b})
in both of the reductions.
We adopt the ``aggressive'' implementation for all datasets  with 
LOCI geometric parameters of $W$ = 5, $N_A$ = 300, $g$ = 1, $N_{\delta}$ = 0.5, and $dr$ = 2
following the definitions in  \citet{Lafreniere:2007bg}.

The SNR for each point source is calculated using aperture photometry with a 5-pix aperture radius.  
The sky background is subtracted from the summed flux centered on the  source using the mean of 100 sky measurements 
at the same angular radius but spanning a range of azimuthal angles surrounding (but not overlapping) the object of interest.  
The standard deviation of these sky values represents the background noise level, 
and the ratio of these two is used to determine the signal to noise of the detection.
Gl 758 B and the two nearby background stars are detected with SNRs between 10 and 51;
full details can be found in Table~\ref{tab:obs}.


PSF subtraction biases astrometry of point sources in processed images as a result of both
over-subtraction and self-subtraction.
To mitigate these effects, we follow the strategy outlined by \citet{Marois:2010hs} of injecting a negative PSF template into the
raw data and iteratively identifying the true position and flux of the sources. 
Three parameters were optimized using the downhill simplex \texttt{AMOEBA} algorithm (\citealt{Nelder:1965tk})--- 
the separation, position angle, and amplitude of the PSF template--- to minimize the resulting rms in a 20-pix aperture radius 
at the location of the point source in the processed image.
Although the host star is visible
behind the coronagraph in the science frames, the mask transmission has historically been difficult to
characterize in detail and may be non-uniform across the face of the occulting spot.
To avoid using a potentially distorted PSF of host star, 
we instead utilize unsaturated PSF templates of other 
stars taken in the same filter (see Table~\ref{tab:obs}).

\begin{figure}
  \vskip -.2 in
  \hskip -.2 in
  \resizebox{3.7in}{!}{\includegraphics{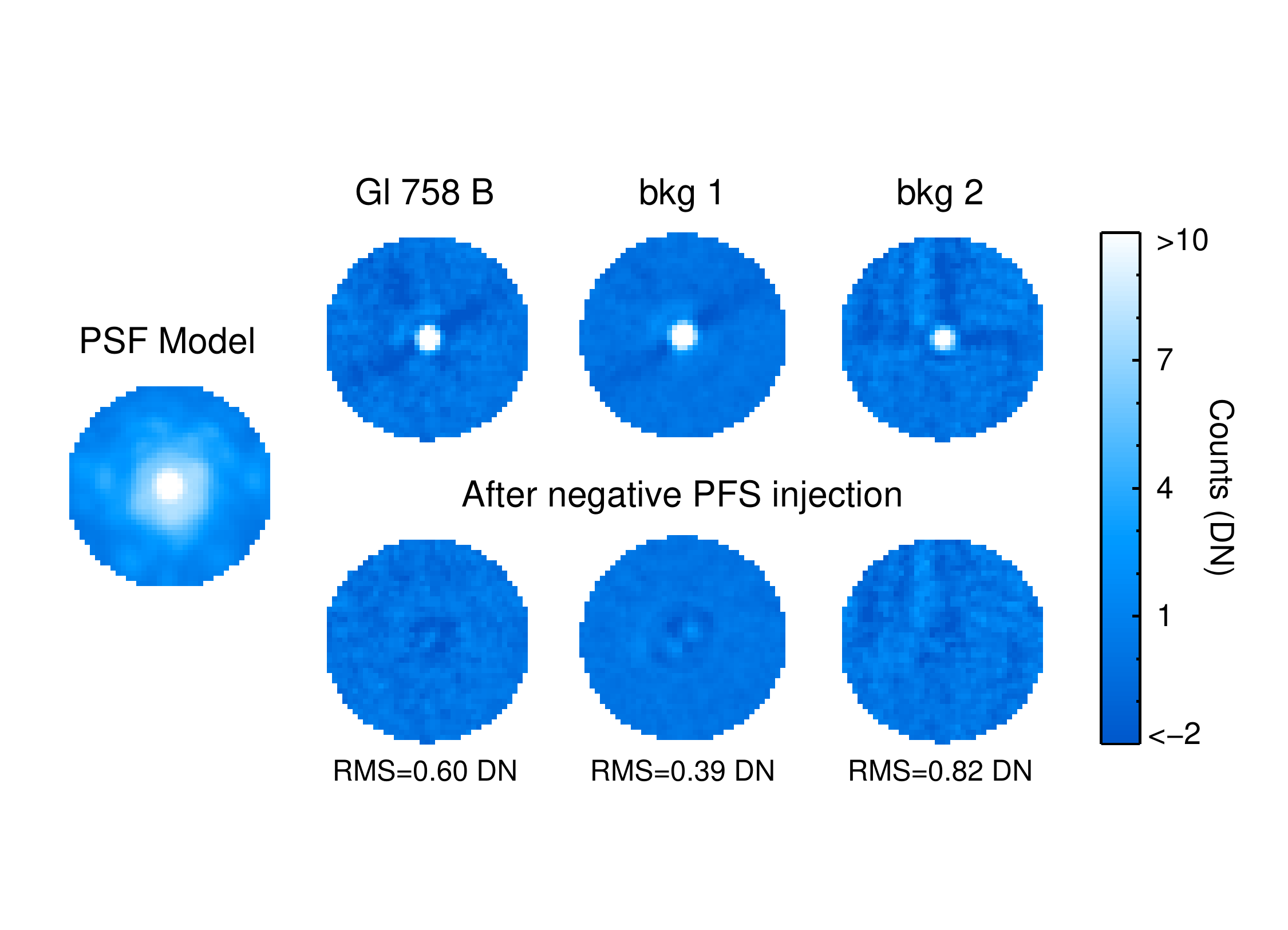}} 
  \vskip -.2 in
  \caption{Example of our negative PSF injection implemented in our 2016 data set 
  to measure unbiased astrometry of point sources.
  On the left is our $H$-band PSF template, in this case created using unsaturated frames 
  of PYC~J11519+0731 taken in May 2012 from \citet{Bowler:2015ch}.  A negative version of this
  model is injected into raw frames, fully processed with LOCI, then iteratively adjusted in position and amplitude 
  to minimize the noise in the final PSF-subtracted image in a 20-pix aperture.  
  Residuals for Gl 758 B, bkg1, and bk2 are shown 
  in the bottom three panels.
     \label{fig:psfinjection} } 
\end{figure}

Results of the negative injection are shown in Figure~\ref{fig:psfinjection}.  This procedure successfully removes
most of the flux and over-subtracted azimuthal ``wings,'' leaving only slight residual structure that likely originates from
an imperfect PSF template, changing atmospheric conditions and Strehl ratios throughout the sequence (which
is not taken into account in the modeling) and/or slight blurring of the PSF if substantial rotation occurs during 
individual exposures--- something that preferentially affects sources at wider separations.


\subsection{Astrometric Error Budget}{\label{sec:asterrors}}

The astrometric error budget is dominated by measurement errors in the positions of point sources;
uncertainty in the residual optical distortion correction; errors in the plate scale and north orientation of the detector; and azimuthal shear caused by sky rotation within individual frames.

Our strategy for estimating point source measurement uncertainties is inspired by the method described in \citet{Rajan:2017hq}
and is intended to mimic the inverted PSF template procedure we carried out in Section~\ref{sec:astrometry}.
For Gl 758 B, bkg1, and bkg2, we inject a positive PSF template into the raw images at the same separation and amplitude 
as the objects of interest but a different position angle.  
We then iteratively inject an inverted PSF template of a different star into the raw data and perform local PSF subtraction with LOCI, then use \texttt{AMOEBA} to identify the optimal position and amplitude that minimizes the local noise.  Each injection/recovery provides an estimate of the systematic difference in the separation, P.A., and amplitude of the injected (positive) object compared with
what was recovered with the inverted (negative) PSF.  
This process is repeated ten times at equally-spaced position angles for each object, and the average positional differences ($\sigma_{\bar{\rho}, meas}$ and $\sigma_{\bar{\theta}, meas}$), are adopted as estimates of the positional measurement errors.

After correcting images for optical distortion effects, there remain small 
residual systematic positional uncertainties, $\sigma_d$, of about 1 mas 
which limits the achievable astrometric accuracy across the detector 
(\citealt{Yelda:2010ig}; \citealt{Service:2016gk}).
Here we adopt one $\sigma_d$ term associated with the host star and one for the companion.
In addition, the NIRC2 plate scale, $s$, and its associated uncertainty, $\sigma_s$, are taken into account
and vary slightly between pre- and post- pupil realignment 
(9.952 $\pm$ 0.002 mas pix$^{-1}$ from \citealt{Yelda:2010ig}; 
9.971 $\pm$ 0.004 mas pix$^{-1}$ from \citealt{Service:2016gk}).

The final separation measurement in mas is 

\begin{equation}{\label{eqn:binsep}}
\rho = s \bar{\rho}_{meas}  ~\pm~   s \bar{\rho}_{meas} \left( \left(\frac{\sigma_{s}}{s}\right)^2 + \left(\frac{\sigma_{\bar{\rho}, tot}}{\bar{\rho}_{meas}}  \right)^2  \right)^{1/2},
\end{equation}

\noindent where $\sigma_{\bar{\rho}, tot}$ is the combined uncertainty from our injection-recovery exercise and 
from the imperfect distortion correction:

\begin{equation}
\sigma_{\bar{\rho}, tot}^2 = \sigma_{\bar{\rho}, meas}^2 + 2 \sigma_{d}^2.
\end{equation}

The P.A. is determined as follows:

\begin{equation}{\label{eqn:binpa}}
\theta = \bar{\theta}_{meas} - \theta_{North} + \theta_{shear}/2,
\end{equation}

\noindent where $\theta_{North}$ is the rotational offset required to align the NIRC2 detector columns with North on the sky: 0.252 $\pm$ 0.009$^{\circ}$ for the \citet{Yelda:2010ig} distortion solution, and 0.262 $\pm$ 0.002$^{\circ}$ for the \citet{Service:2016gk} distortion solution.\footnote{The position angle 
of celestial North with respect to the +$y$ axis for NIRC2 images
taken in vertical angle (pupil tracking) mode with the narrow camera can be found using FITS header 
keywords: \texttt{PARANG} + \texttt{ROTPOSN} -- \texttt{INSTANGL} -- $\theta_{North}$.
Note that $\theta_{North}$ is \emph{subtracted} from the other terms (J. Lu, M. Service, private communication, 2017).}
$\theta_{shear}$ is the shear (blurring) per individual frame.  To account for this, each frame is 
de-rotated to the midpoint P.A. of the exposure after PSF subtraction, and prior to coaddition of the sequence.

The error in the P.A., $\sigma_{\theta}$, includes the injection-recovery measurement uncertainty ($\sigma_{\theta, meas}$), 
uncertainty in the north alignment ($\sigma_{\theta, North}$), residual positional errors after applying
the distortion solution ($\sigma_{\theta, d}$), and the systematic error from shearing
of point sources from sky rotation within each frame ($\sigma_{\theta, shear}$):

\begin{equation}{\label{eqn:binpa}}
\sigma_{\theta} = \left( \sigma_{\theta, meas}^2 + \sigma_{\theta, North}^2 + \sigma_{\theta, d}^2 +  \sigma_{\theta, shear}^2 \right)^{1/2}.
\end{equation}

\noindent The residual positional distortion errors are about 1~mas, so here we 
approximate $\sigma_{\theta, d}$ as $\approx$1 mas/$\rho$.  The dominant term in the P.A.  
error budget is the shear per frame, which varies among individual frames and across observation epochs.  
For this work we conservatively adopt half the average shear for each epoch:
0$\fdg$30, 0$\fdg$21, 0$\fdg$38, and 0$\fdg$32 for our 2010, 2013, 2016, and 2017 epochs.
Our final astrometry of Gl 758 B and the two background sources are listed in Table~\ref{tab:obs}.
Note that the resulting astrometry does not appear to be significantly sensitive to changes 
in the LOCI parameters used for PSF subtraction based on the same astrometric analysis
with $N_{\delta}$ set to 1.5.


\begin{figure}
  \vskip -.4 in
  \hskip -1.5 in
  \resizebox{7.1in}{!}{\includegraphics{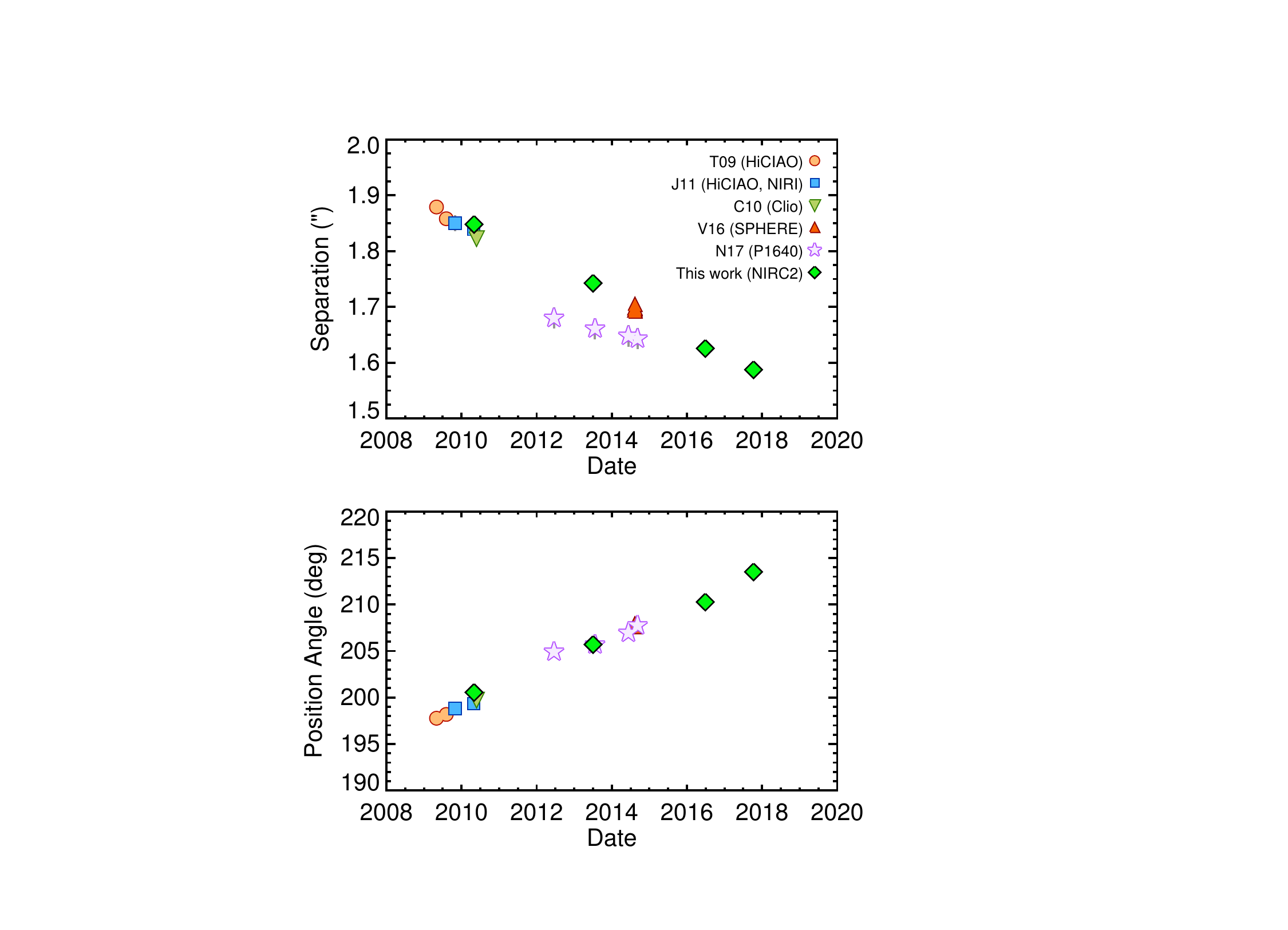}} 
  \vskip -.4 in
  \caption{ Relative astrometry of Gl 758 B.  
  The companion is approaching Gl 758 at a rate of $\approx$34 mas yr$^{-1}$ (top panel) and orbiting 
  in a counterclockwise direction on the sky by 1$\fdg$86 yr$^{-1}$ (bottom panel).  Astrometry are from\
  \citet[T09]{Thalmann:2009ca}, \citet[J11]{Janson:2011dh}, \citet[C10]{Currie:2010ju}, 
  \citet[V16]{Vigan:2016gq}, and \citet[N10]{Nilsson:2017hm}.
  Diamonds indicate our new epochs taken with Keck/NIRC2.
     \label{fig:astrometry} } 
\end{figure}

\subsection{Comparison to Published Astrometry}{\label{sec:pubastrometry}}

Gl 758 B has been observed by many other telescopes and instruments over the past decade (Figure~\ref{fig:astrometry}). 
The companion displays clear orbital motion; its separation has contracted from 1$\farcs$88 in 2009 
to 1$\farcs$59 with our latest epoch from NIRC2 in 2017, and has moved by $\approx$16$^{\circ}$ in P.A. during that time.
Our astrometry of Gl 758 B is broadly consistent with published values, although
the separations from \citet{Nilsson:2017hm} are significantly smaller than our measurements and those
of \citet{Vigan:2016gq} taken over the same time period.  
For example, our NIRC2 observations from 2013 were taken within three weeks of 
the 21 July 2013 dataset obtained by Nilsson et al., but these two separation measurements are discrepant at the 4.3-$\sigma$
level.
However, the P.A. measurements from Nilsson et al. are in much better agreement.


\begin{figure*}
  \vskip 0 in
  \hskip .8 in
  \resizebox{5in}{!}{\includegraphics{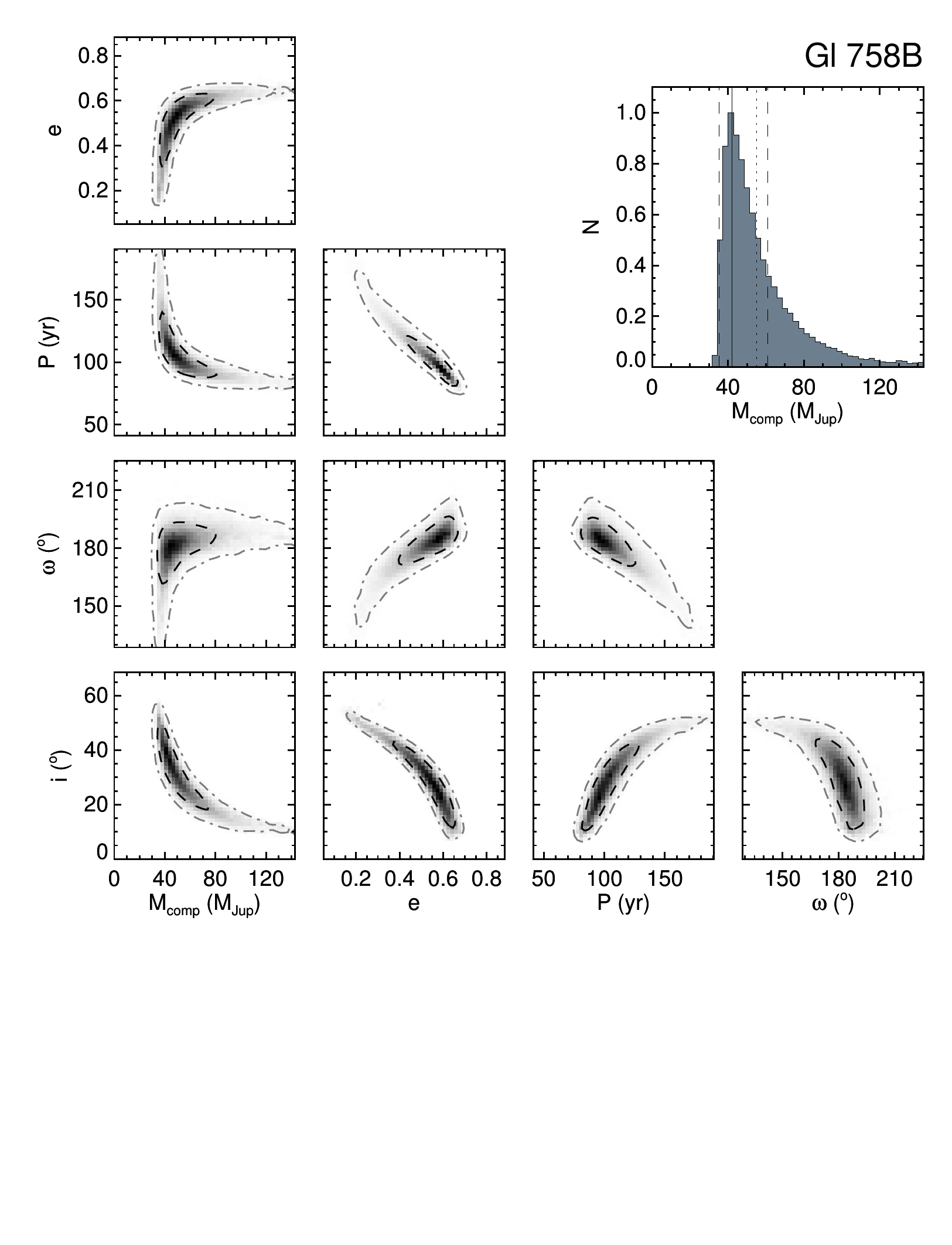}} 
  \vskip -1.7 in
  \caption{Posterior distributions of orbital parameters from our MCMC analysis.
Grayscale images show the relation between the companion mass
($M_{\rm comp}$) and the several of the most correlated parameters: inclination,
eccentricity, period, and the argument of periastron.
Contours indicate the regions
containing 68.3\% (black dashed lines) and 95.4\% (gray dash-dotted
lines). The histogram (top right) shows the marginalized posterior of
$M_{\rm comp}$ with the mode (solid line), best-fit (dotted line), and
68.3\% (1$\sigma$) interval (dashed lines) indicated.
 \label{fig:corner} } 
\end{figure*}

\section{Orbit and Dynamical Mass of Gl 758 B}{\label{sec:orbit}}

\subsection{Is the Acceleration Caused by Gl 758 B?}{\label{sec:acceleration}}

Before carrying out a detailed joint orbit fit of the RVs and astrometry, we first demonstrate here that 
the observed acceleration of Gl 758 is consistent with and likely to be caused 
by the companion Gl 758 B.
The minimum mass of an imaged companion needed to 
produce an observed instantaneous acceleration $\dot{v}_r$ is

\begin{equation}
M \approx 0.0145   \left( \frac{d}{\mathrm{pc}} \ \frac{\rho}{''}   \right) ^2 \Big|  \frac{\dot{v}_r}{\mathrm{m} \ \mathrm{s^{-1}} \ \mathrm{yr^{-1}}} \Big|  \Mjup, 
\end{equation}

\noindent where $d$ is the distance to the system in pc and $\rho$ is the projected separation in arcseconds
(\citealt{Torres:1999gc}; \citealt{Liu:2002fx}).  
Note that the generalized form of this equation includes information about the orbital elements of the system
in the form of a multiplicative constant, the minimum value of which ($\approx$2.6) is included here in the prefactor.
The measured range of accelerations 
and angular separations of Gl 758 B imply a
corresponding mass range of $\approx$20--50~\Mjup.
Are these reasonable values for Gl 758 B?
Evolutionary models from \citet{Saumon:2008im} suggest that a brown dwarf
with those masses should have effective temperatures between 
about 430--1400~K for ages of 1--6~Gyr; this is in good agreement
with the inferred effective temperatures of 600--750 K for Gl 758 B from multi-band imaging
and spectroscopy (\citealt{Vigan:2016gq}; \citealt{Nilsson:2017hm}).
Based on this consistency and the fact that the companion's orbital period must be 
much longer than the time baseline of the RV observations ($\gg$20~yr),
we conclude that Gl 758 B is likely the culprit of the acceleration.


\begin{figure*}
  \vskip 0 in
  \hskip 0.8 in
  \resizebox{5in}{!}{\includegraphics{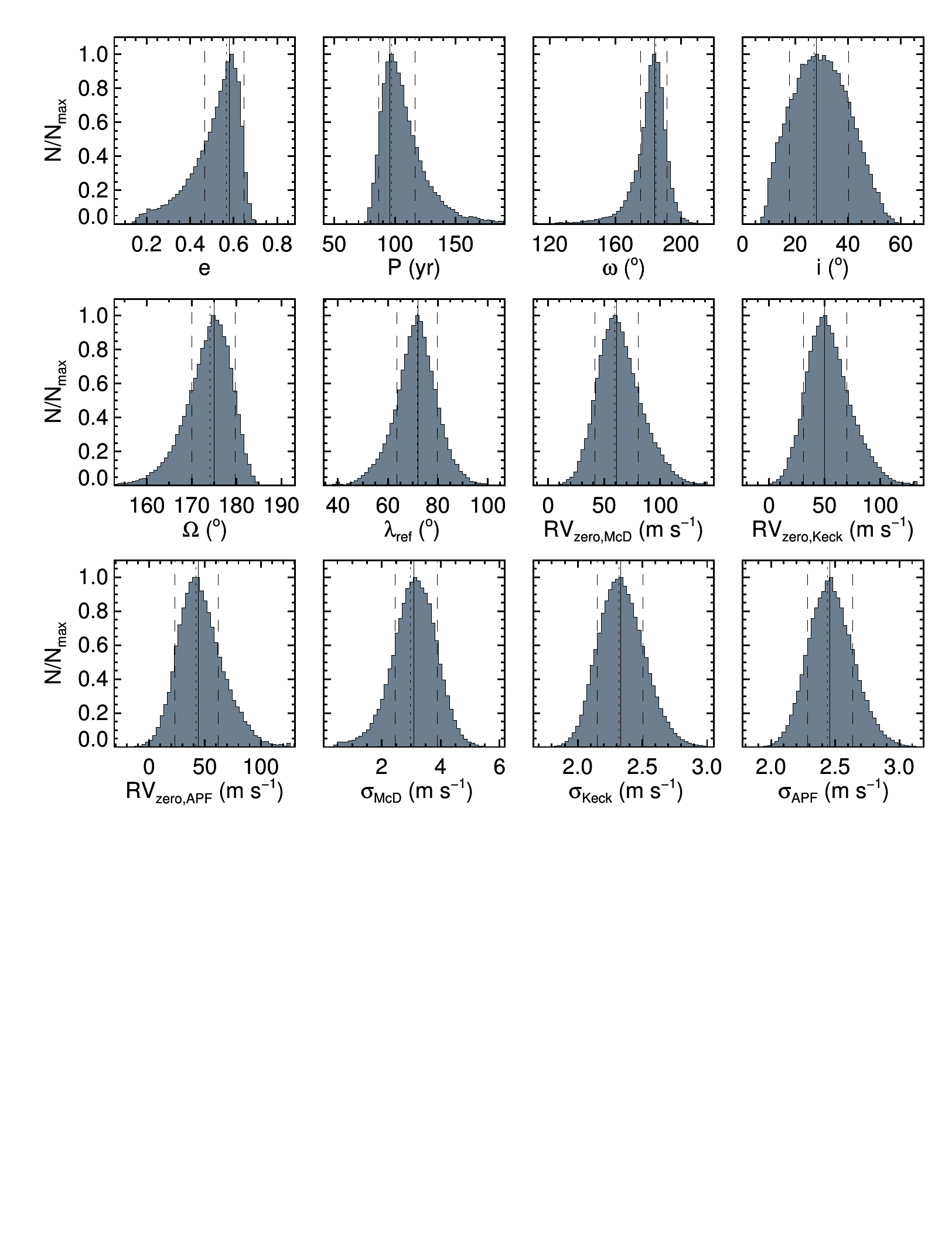}} 
  \vskip -2.3 in
  \caption{Marginalized posteriors of fitted parameters from our MCMC analysis.
Vertical solid lines show the modes, dotted lines show the best-fit
values, and dashed lines show the 68.3\% (1$\sigma$) intervals.
\label{fig:hist} } 
\end{figure*}

\subsection{Joint RV and Astrometric Orbit Analysis}{\label{sec:jointorbit}}

We performed a joint orbit analysis of the three RV data sets and
astrometry via a Markov Chain Monte Carlo (MCMC) algorithm. 
For this analysis we only use our NIRC2 astrometry 
to avoid systematic errors that may be present in previously published astrometry
caused by multiple instruments and PSF subtraction strategies.
We used the parallel-tempering (PT) ensemble sampler in \texttt{emcee~v2.1.0}
(\citealt{ForemanMackey:2013io}) that is based on the algorithm described
by \citet{Earl:2005hv}. 30 temperatures were adopted, of which only
the coldest chain describes the posterior, together with 100 walkers to sample
our 15-parameter model. Six of those parameters describe the orbit:
semimajor axis ($a$), inclination ($i$), P.A. of the ascending node
($\Omega$), mean longitude ($\lambda_{\rm ref}$) at a reference epoch ($t_{\rm ref}$) of
2455197.5~JD, and finally eccentricity ($e$) and
the argument of periastron ($\omega$, for the host star) parameterized as
$\sqrt{e}\sin{\omega}$ and $\sqrt{e}\cos{\omega}$, which avoids the Lucy-Sweeney 
bias toward non-zero eccentricities and imposes a uniform prior on eccentricity. 
We assumed a
log-flat prior on $a$, randomly distributed viewing angles for $i$
(i.e., a prior of $\sin{i}$), and uniform priors for the other orbit
parameters. The next three parameters used in the fit were the
parallax ($\pi$) and mass of the host star ($M_{\rm host}$) and the
mass of the companion ($M_{\rm comp}$). We assumed priors of
$63.45\pm0.35$\,mas on the parallax (\citealt{vanLeeuwen:2007dc}),
$0.97\pm0.02$\,\Msun\ for $M_{\rm host}$ (\citealt{Vigan:2016gq}),
and a log-flat prior for $M_{\rm comp}$, which is motivated by the
broad range of potential masses for the companion spanning $\approx$10~\Mjup \
(if the system is younger than expected) to over 100~\Mjup \ (if the companion 
is an unresolved binary).
The remaining six parameters
are the zero points ($\Delta_{\rm zero, {\rm McD}}$, $\Delta_{\rm zero, {\rm
    Keck}}$, $\Delta_{\rm zero, {\rm APF}}$) and jitters ($\sigma_{\rm jit,
  Keck}$, $\sigma_{\rm jit, APF}$, $\sigma_{\rm jit, McD}$) for the
three RV data sets. The zero points are simply the offsets needed to
bring the sets of relative RVs into accord with the orbit model, and the
jitter terms account for small random and systematic RV epoch-to-epoch measurement errors from the star and the instrument
not captured in the quoted relative RV uncertainties. 
We assumed uniform priors for the
zero points and log-flat priors for the jitters.  
These are summarized in Table~\ref{tab:mcmc}, and our 
complete likelihood function is as follows:

\begin{equation}
\begin{split}
  \ln(\mathcal{L}) = -0.5 \left( \sum_{k=1}^{N_{\rm ast}} \left(\frac{\rho_k-\rho(t_k)}{\sigma_{\rho, k}}\right)^2  + 
    \sum_{k=1}^{N_{\rm ast}} \left(\frac{\theta_k-\theta(t_k)}{\sigma_{\theta, k}}\right)^2 \right. \\
    \left. + \sum_{j=1}^{N_{\rm inst}}\sum_{k=1}^{N_{\rm RV}} \frac{\left({\rm RV}_{{\rm rel},k}+\Delta_{{\rm zero},j}-{\rm RV}(t_k) \right)^2}{\sigma_{{\rm RV}, k}^2 + \sigma_{{\rm jit}, j}^2} \right) \\
+ \ln(\sin(i)) - \ln(a) - \ln(M_{\rm sec}) - 0.5\left( \frac{\pi-63.45\,{\rm mas}}{0.35\,{\rm mas}} \right)^2 
\end{split}
\end{equation}

We are able to fit for the RV jitter because of the
numerous independent data points that sample its orbit in each data
set. In other words, there are many degrees of freedom in the RV model
of the host star. In contrast, with only four epochs of companion
astrometry we do not have the same ability to fit for additional
astrometric errors. Therefore, we performed an initial orbit fit using
the nominal astrometric errors and examined the residuals. The rms of
the separation measurements was 4.3\,mas about this initial best-fit
orbit and $\chi^2=14.6$, while for the P.A.s the rms was 0.34\degr\
with $\chi^2=3.0$. As a point comparison, when we simply fit a line to
separation and P.A. as a function of time, we found similar rms values
of of 4.7\,mas ($\chi^2=17.5$) and 0.45\degr\ ($\chi^2$=4.9). Given
that the RVs constrain some of the same orbit parameters that are
relevant to the astrometric fit, it is not obvious what number of
degrees of freedom is correct to assume here. If we assume two degrees
of freedom in each, then $p(\chi^2)=0.0007$ for the separations and
$p(\chi^2)=0.22$ for the P.A.s. Ultimately we add 4.3\,mas in quadrature to our
separation measurements, resulting in effective errors of $\approx$5\,mas at every epoch. 
We do not add any additional error to our P.A. uncertainties as 
these are already substantially larger ($\approx$0$\fdg$3, or $\approx$10~mas) and the
$\chi^2$ value is not unreasonable. 
The source of the 4--5\,mas epoch-to-epoch
uncertainties in our separation measurements is not known, so it may
represent a fundamental floor to astrometry derived from ADI sequences
with the NIRC2 coronagraph.

\begin{deluxetable*}{lcccc}
\renewcommand\arraystretch{0.9}
\tabletypesize{\small}
\tablewidth{0pt}
\tablehead{
\colhead{Property}              &
\colhead{Mode   $\pm$1$\sigma$} &
\colhead{Best fit}              &
\colhead{95.4\% c.i.}           &
\colhead{Prior}                 }
\tablecaption{MCMC Posteriors for the Orbit of Gl~758B \label{tab:mcmc}}
\startdata
\multicolumn{5}{c}{Fitted parameters} \\[1pt]
\cline{1-5}
\multicolumn{5}{c}{} \\[-5pt]
Companion mass $M_{\rm comp}$ (\Mjup)                                       & $42_{-7}^{+19}$                & 55            &           33, 106          & $1/M$ (log-flat)                           \\[3pt]
Host-star mass $M_{\rm host}$ (\Msun)                                       & $0.967_{-0.018}^{+0.022}$      & 0.969         &        0.929, 1.009        & $0.970\pm0.020$\,\Msun\ (Gaussian)         \\[3pt]
Parallax (mas)                                                              & $63.39_{-0.32}^{+0.37}$        & 63.51         &        62.73, 64.13        & $63.45\pm0.35$\,mas (Gaussian)             \\[3pt]
 Semimajor axis $\alpha$ (mas)                                               & $1340_{-80}^{+170}$            & 1350          &         1210, 1820         & $1/\alpha$ (log-flat)                      \\[3pt]
 Inclination $i$ (\degr)                                                     & $28_{-10}^{+12}$               & 27            &           10, 49           & $\sin(i)$, $0\degr < i < 180\degr$         \\[3pt]
$\sqrt{e}\sin{\omega}$                                                      & $-0.05_{-0.09}^{+0.11}$        & $-$0.06       &      $-$0.26, 0.24         & uniform                                    \\[3pt]
$\sqrt{e}\cos{\omega}$                                                      & $-0.76_{-0.03}^{+0.08}$        & $-$0.75       &      $-$0.82, $-$0.47      & uniform                                    \\[3pt]
Mean longitude at $t_{\rm ref}=2455197.5$~JD, $\lambda_{\rm ref}$ (\degr)   & $72\pm8$                       & 72            &           52, 89           & uniform                                    \\[3pt]
PA of the ascending node $\Omega$ (\degr)                                   & $175\pm5$                      & 174           &          163, 183          & uniform                                    \\[3pt]
McDonald RV zero point (m\,s$^{-1}$)                                        & $61\pm19$                      & 59            &           27, 108          & uniform                                    \\[3pt]
Keck RV zero point (m\,s$^{-1}$)                                            & $50_{-19}^{+20}$               & 50            &           17, 98           & uniform                                    \\[3pt]
APF RV zero point (m\,s$^{-1}$)                                             & $44_{-21}^{+18}$               & 42            &            9, 90           & uniform                                    \\[3pt]
McDonald RV jitter $\sigma_{\rm McD}$ (m\,s$^{-1}$)                         & $3.1_{-0.6}^{+0.8}$            & 3.0           &          1.6, 4.6          & $1/\sigma$ (log-flat)                      \\[3pt]
Keck RV jitter $\sigma_{\rm Keck}$ (m\,s$^{-1}$)                            & $2.33_{-0.18}^{+0.17}$         & 2.32          &         2.00, 2.70         & $1/\sigma$ (log-flat)                      \\[3pt]
APF RV jitter $\sigma_{\rm APF}$ (m\,s$^{-1}$)                              & $2.46_{-0.17}^{+0.18}$         & 2.44          &         2.13, 2.84         & $1/\sigma$ (log-flat)                      \\[3pt]
\cline{1-5}
\multicolumn{5}{c}{} \\[-5pt]
\multicolumn{5}{c}{Computed properties} \\[1pt]
\cline{1-5}
\multicolumn{5}{c}{} \\[-5pt]
Orbital period $P$ (yr)                                                     & $96_{-9}^{+21}$                & 97            &           79, 153          & \nodata                                    \\[3pt]
Semimajor axis $a$ (AU)                                                     & $21.1_{-1.3}^{+2.7}$           & 21.3          &         18.9, 28.7         & \nodata                                    \\[3pt]
Eccentricity $e$                                                            & $0.58_{-0.11}^{+0.07}$         & 0.57          &         0.26, 0.67         & \nodata                                    \\[3pt]
Argument of periastron $\omega$ (\degr)                                     & $184_{-9}^{+8}$                & 184           &          153, 201          & \nodata                                    \\[3pt]
Time of periastron $T_0=t_{\rm ref}-P\frac{\lambda-\omega}{360\degr}$ (JD)  & $2465800_{-800}^{+2000}$       & 2466300       &      2464800, 2470500      & \nodata                                    \\[3pt]
Mass ratio $q = M_{\rm comp}/M_{\rm host}$                                  & $0.042_{-0.008}^{+0.018}$      & 0.054         &        0.032, 0.105        & \nodata                                    \\[3pt]
\enddata
\end{deluxetable*}


\begin{figure*}
  \vskip 0 in
  \hskip 0.4 in
   \includegraphics[width=3.in]{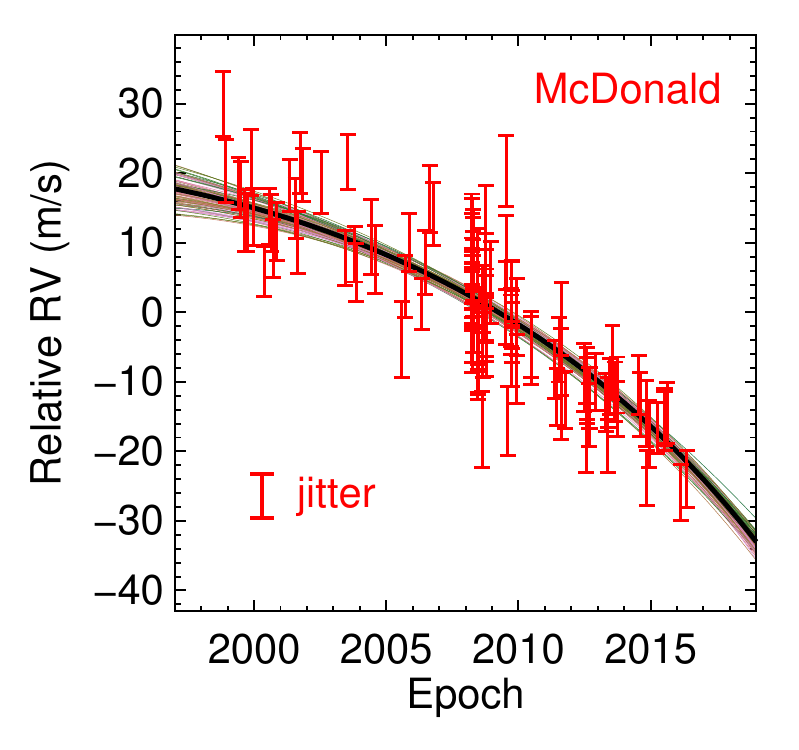}\includegraphics[width=3.in]{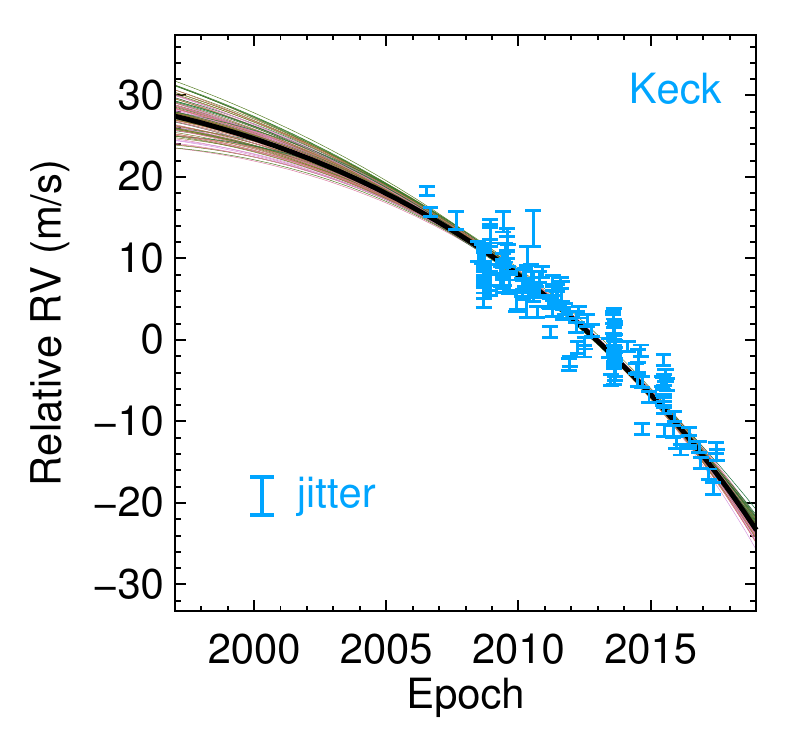}
  \vskip 0 in
  \hskip .4 in
   \includegraphics[width=3.in]{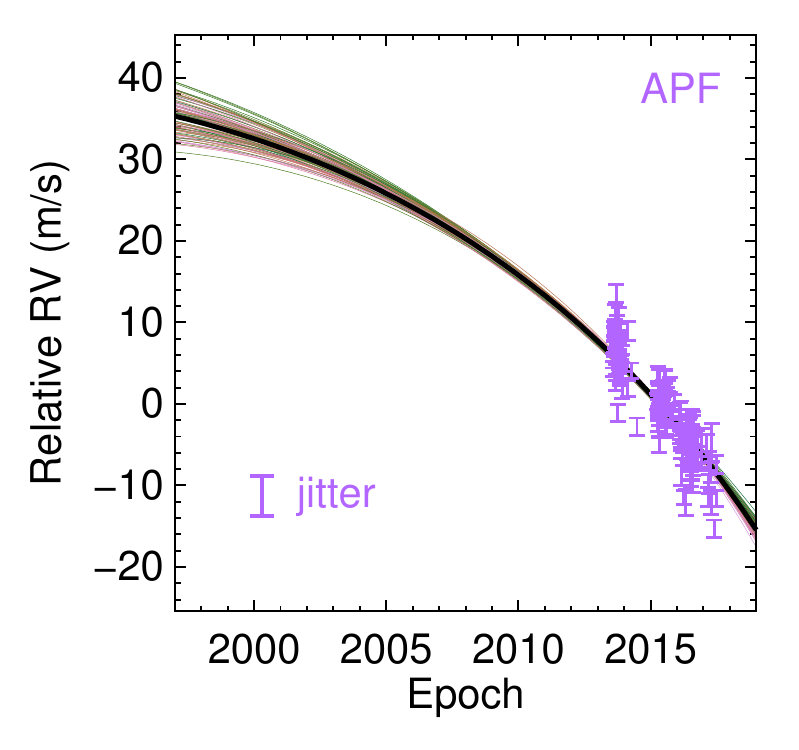}\includegraphics[width=4.3in]{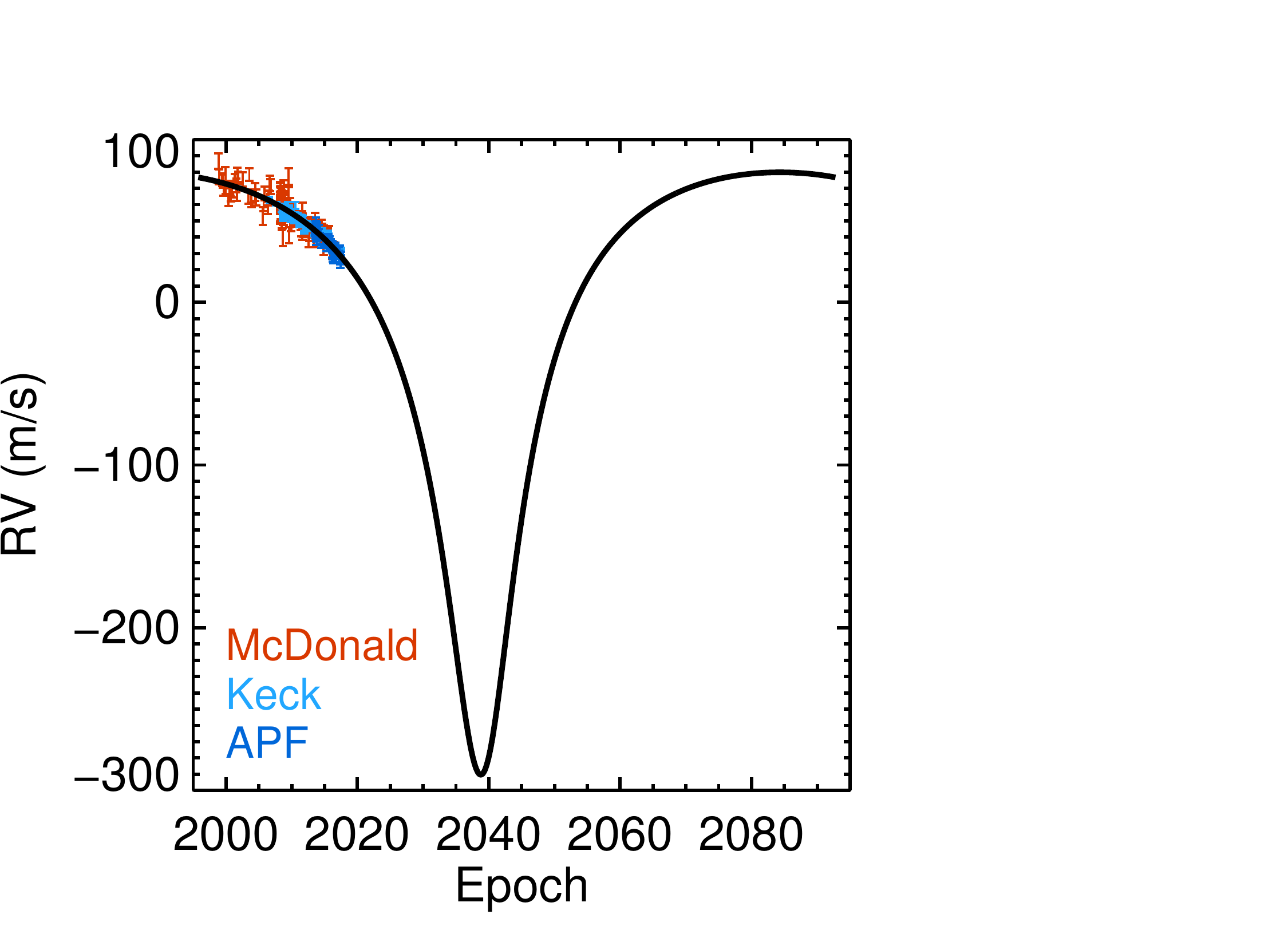}
  \vskip -.2 in
  \caption{Relative RV measurements of Gl~758 from McDonald (top left), Keck
(top right), and APF (bottom left). Randomly drawn orbit solutions from our
MCMC posterior are displayed as thin colored lines, coded by the
companion mass from low (pink) to high (green). Each RV data set has 
its own RV zero point associated with each orbit solution, allowing
the absolute, barycentric RVs predicted from the orbit to be
plotted as relative RVs here. The best-fit orbit solution is shown as
a thick black line, and the error bar in the lower left of each plot
shows the best-fit jitter (i.e., the additional RV error that is added
in quadrature to the displayed measurements during our MCMC analysis).
The best-fit orbit solution (black line) for the barycentric velocity
of the primary over time. Each RV data set is shown with its best-fit
zero point added to bring the relative RVs into the barycentric frame.
Jitter has not been added to the plotted error bars. The RV
measurements jointly show a nonlinear trend, indicating that the
acceleration of the host star is changing with time.  The bottom-right panel
shows the RVs relative to the best-fit orbit spanning a complete orbital cycle (97 yr).
   \label{fig:rvfit} } 
\end{figure*}

The initial state of the PT-MCMC sampler was determined using a Monte
Carlo rejection sampling analysis similar to the method used in
\citet{Dupuy:2016dh}. First, $2\times10^6$ randomly distributed
orbital periods ($10^4\,{\rm d} < P < 10^7$\,d), eccentricities, and
times of periastron passage ($T_0$) were drawn. Using the formalism of
\citet{Lucy:2014kr}, we computed the corresponding set of $a$,
$i$, $\omega$, and $\Omega$ that best fit the astrometry for each of
these trials. For each trial, we computed the $\chi^2_{\rm ast}$ of
the trial orbit's predicted astrometry and our measured astrometry. To
incorporate the RVs, we assumed at this stage that each data set could
be represented as a simple linear trend with time. For each orbit
trial, we computed the instantaneous slope of the host-star RV at the
mean epoch for each RV data set. (In our actual PT-MCMC runs, we fit
all the individual relative RV measurements directly.) Because each
trial has $a$ and $P$ independent of assumptions about mass, each
trial also effectively samples an associated total mass through $M_{\rm tot} \propto a^3/P^2$. We computed the mass
ratio that would best bring each orbit's RV slopes into agreement with
the measured slopes and then computed the RV zero points needed to
bring the measured relative RVs into agreement. Because there are
multiple RV slopes to reproduce, the agreement is not perfect for a
given orbit trial, and we computed the $\chi^2_{\rm RV}$ of the trial
orbit's predicted RV slopes and our measured RV slopes. Finally, we
computed the $\chi^2_{\rm mass}$ of the trial orbit's predicted
host-star mass with the estimated mass of 0.97~\Msun \ from
\citet{Vigan:2016gq}, with an inflated uncertainty of
$\pm0.20$\,\Msun\ to allow us to perform orbit fits with no mass
prior as well. We combined these constraints into
$\chi^2_{\rm tot} = \chi^2_{\rm ast} + \chi^2_{\rm RV} +\chi^2_{\rm
  mass}$, computed rejection probabilities of $p_{\rm
  rej} = 1 - \exp(-(\chi^2_{\rm tot}-{\rm min}(\chi^2_{\rm tot}))/2)$,
and then drew random samples to pass on based on $p_{\rm rej} > \mathcal{U}(0,1)$,
where $\mathcal{U}(0,1)$ was a uniformly distributed, randomly drawn number ranging
from 0 to 1.


\begin{figure*}
  \vskip 0. in
  \hskip .1 in
  \includegraphics[width=3.2in]{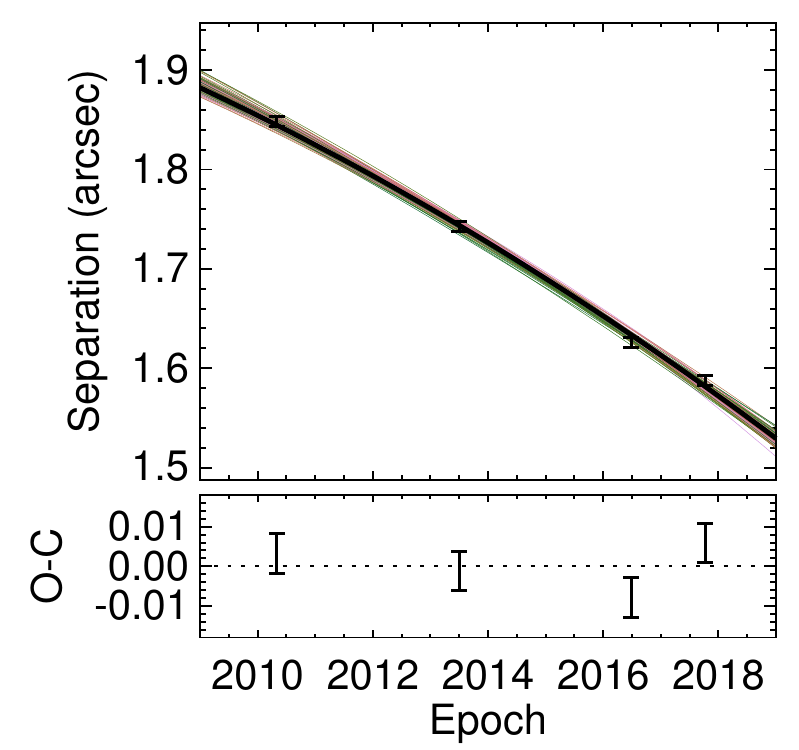}\includegraphics[width=3.2in]{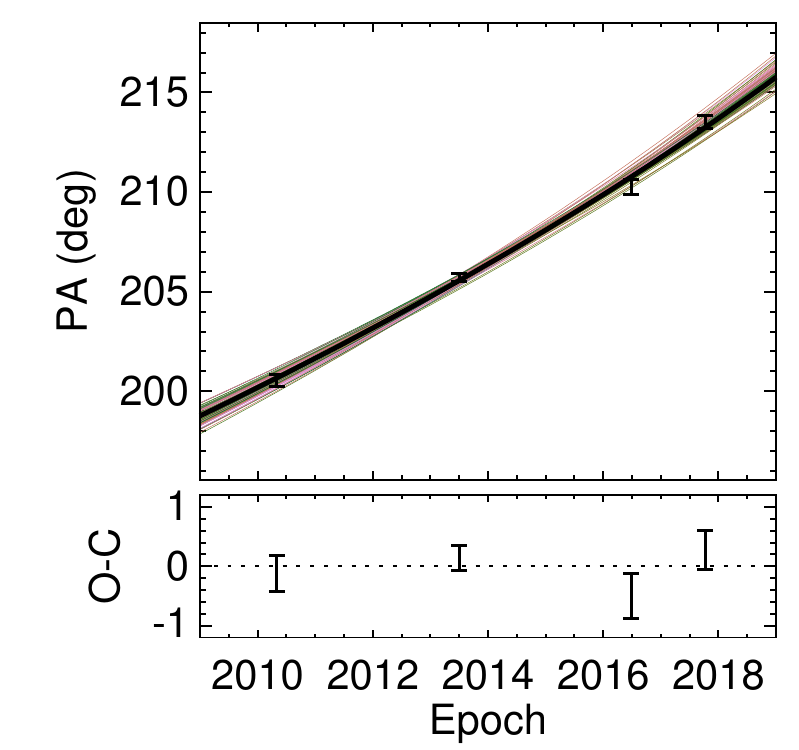}
  \vskip 0 in
  \caption{Separation (left) and P.A. (right) measurements of Gl~758~B relative to
the primary. Randomly drawn orbit solutions from our MCMC posterior
are displayed as thin colored lines, coded by the companion mass from
low (pink) to high (green). The best-fit orbit solution is shown as a
thick black line.
\label{fig:orbitfit} } 
\end{figure*}


\begin{figure*}
  \vskip -.1 in
  \hskip .1 in
  \includegraphics[width=3.2in]{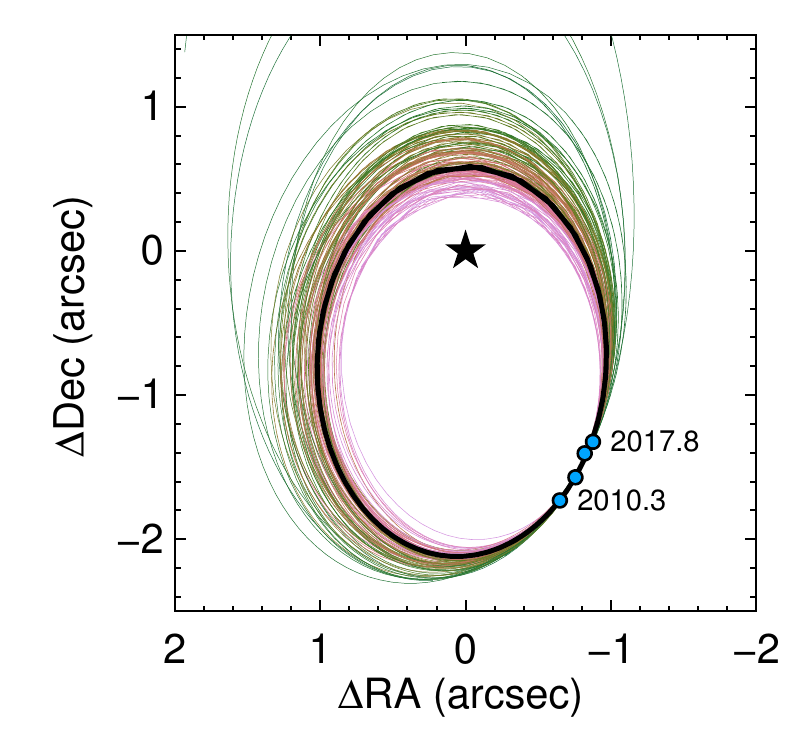}\includegraphics[width=3.2in]{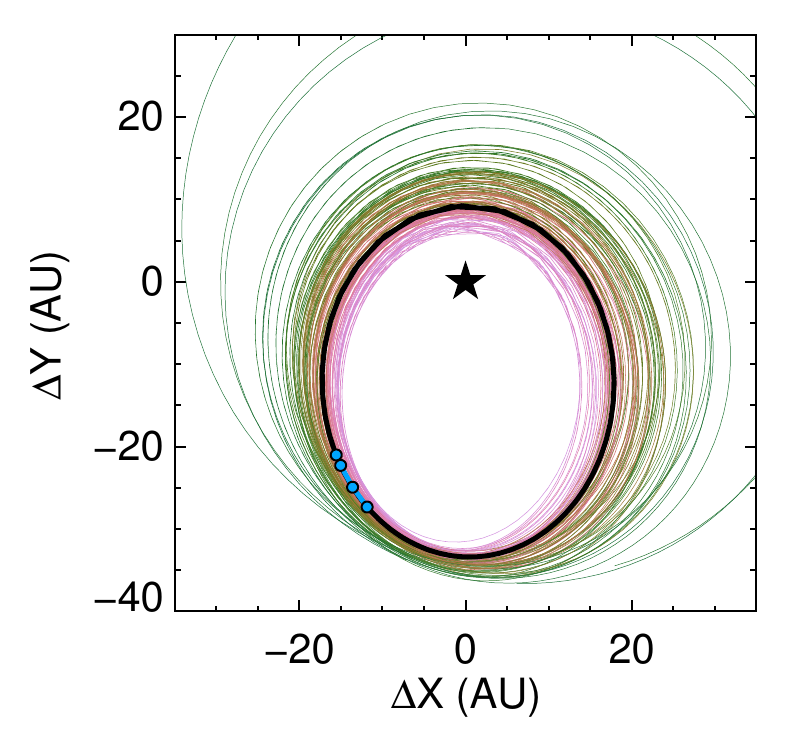}
  \vskip 0 in
  \caption{Left: Astrometry of Gl~758~B (blue circles) relative to the primary
(black star) shown alongside randomly drawn orbit solutions from our
MCMC posterior. Right: The same orbit solutions viewed face-on (i.e.,
$i$ set to zero) with the epochs of our astrometry marked in blue. The
best-fit orbit solution is shown as a thick black line. The randomly
drawn orbits are color coded according to the companion mass, from low
mass (pink) to high mass (green). Higher companion mass is strongly
correlated with smaller, more eccentric, and closer to face-on orbits.
\label{fig:radec} } 
\end{figure*}

In our PT-MCMC analysis, we experimented with different chain lengths
and found that after $\sim$10$^5$ steps our 100-walker chains had
clearly stabilized in the mean and rms of the posterior for each of
the parameters. We saved every 20th step of our chains and discarded
the first 50\% of the chain as the burn-in portion, leaving
$2.5\times10^5$ PT-MCMC samples in the cold chain.
Table~\ref{tab:mcmc} lists information on the posterior distributions
of our fitted parameters, as well as parameters that are directly
computed from them. To compute the modes of our distributions we
binned the posterior and found the bin with the most elements. The 1-
and 2-$\sigma$ confidence intervals are computed as the minimum range
in that parameter that contains 68.3\% and 95.4\% of the values,
respectively. The quoted best-fit solution  is the one with the maximum
likelihood, which includes the prior.

Figure~\ref{fig:corner} displays the companion mass posterior and the
most relevant parameter correlations, and Figure~\ref{fig:hist} shows several of the
other marginalized posterior distributions from our fit. As expected, the inclination is
highly correlated with companion mass (i.e., $M_{\rm comp}\sin(i)$ is
well constrained from the RV orbit). The companion mass posterior extends to
high masses ($>$80\,\Mjup) that are likely unphysical assuming that
Gl~758 B is a single object. This high mass tail corresponds to low
inclinations ($i\lesssim20$\degr) that our astrometry cannot rule
out, 
high eccentricities
($e\gtrsim$0.6), and short periods ($P\lesssim$100\,yr).
The companion mass posterior has a sharp lower limit of 30.5~\Mjup \
at the 4-$\sigma$ level.\footnote{Note that this lower limit on the companion mass is 
relatively insensitive to changes in priors.  We also ran our joint fit after removing the
prior on the host star mass; the resulting mode of the companion mass distribution is 43~\Mjup \ 
with a 95\% credible interval of 33--131~\Mjup \ and a lower limit of 31.7~\Mjup (at the 4-$\sigma$ level).}
Figures~\ref{fig:rvfit} and \ref{fig:orbitfit} show our orbit solutions
relative to our RV and astrometric data.
The sky-projected and de-projected solution for Gl 758 is displayed in
Figure~\ref{fig:radec}, where orbits are drawn from the 
MCMC posteriors and are color-coded 
according to the corresponding companion mass from low mass (pink) to
high mass (green).

\section{Results and Discussion}{\label{sec:discussion}}

\subsection{Nature of bkg2}{\label{sec:bkg2}}

\begin{figure}
  \vskip 0 in
  \hskip -.3 in
  \resizebox{3.8in}{!}{\includegraphics{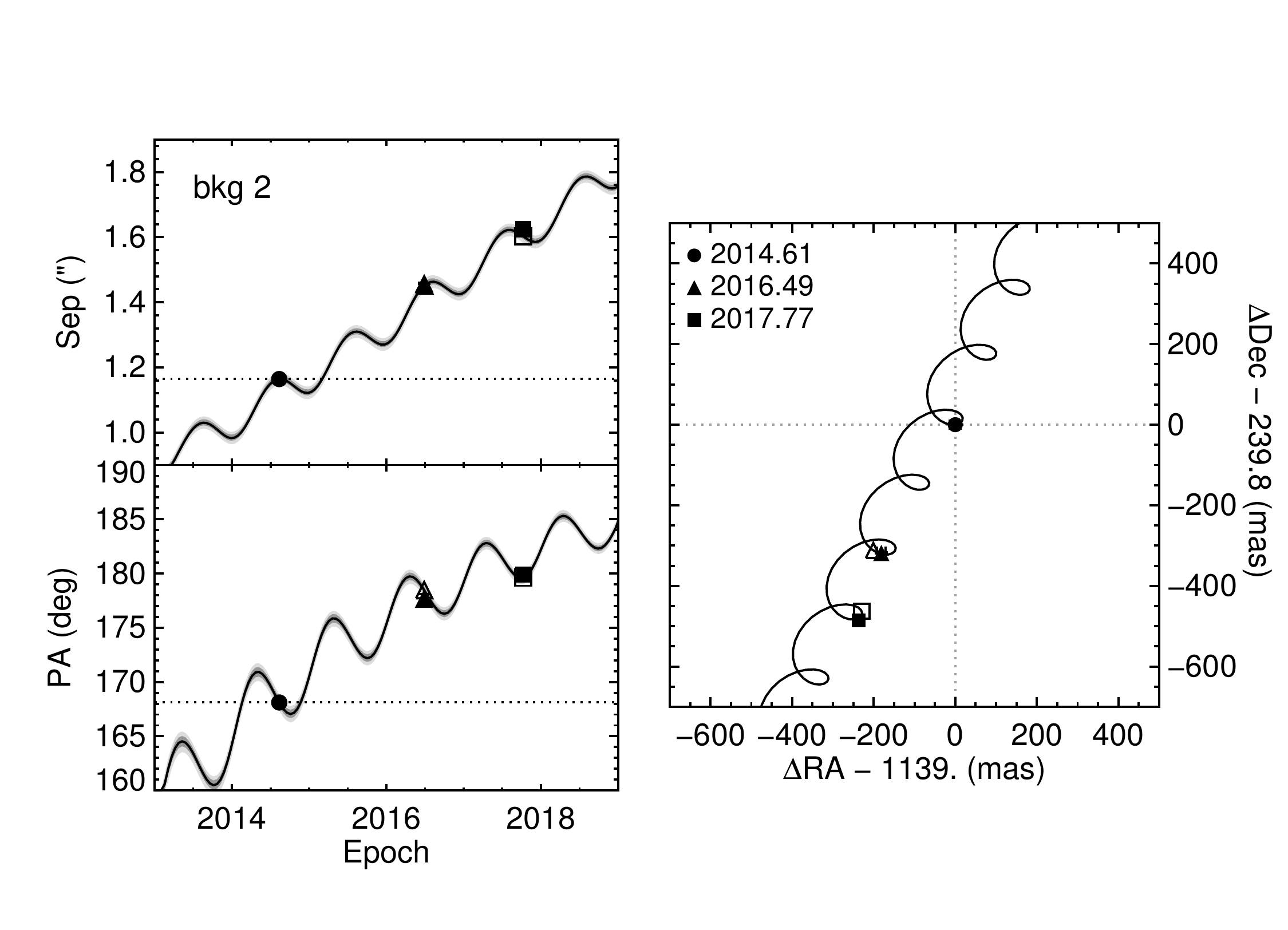}} 
  \vskip -.2 in
  \caption{Astrometry of ``bkg2'' relative to Gl 758.  This object closely follows the expected
  trajectory of a background star (black track with 1- and 2-$\sigma$ uncertainties shown in gray) 
  based upon the proper motion and parallax of Gl 758 in separation (top), P.A. (bottom), and 
  relative position on the sky (right).  Open symbols represent the expected position
  of a background object at the time of our observations; filled symbols are our measured values.
  Uncertainties are generally smaller than the symbol sizes.  The 2014 epoch is from \citet{Vigan:2016gq},
  and our new NIRC2 astrometry of bkg2 were taken in 2016 and 2017.     \label{fig:bkg2} } 
\end{figure}

\citet{Vigan:2016gq} identified a new point source near Gl 758 at a separation of $\approx$1$\farcs$1
based on  observations taken with SPHERE in 2014.
They found that the photometry of this new object  
is broadly consistent with the colors of L dwarfs, raising the possibility 
that this could be a second companion in this system.  
This object is easily recovered in our 2016 and 2017 datasets at wider separations of 1$\farcs$46 and 1$\farcs$62.  
Together with the single-epoch detection from Vigan et al. taken in 2014, this object closely follows the expected motion
for a background star (Figure~\ref{fig:bkg2}) and appears to be unassociated with Gl 758.  


\subsection{Bolometric Luminosity of Gl 758 B}{\label{sec:luminosity}}

Despite extensive efforts to characterize this system with follow-up photometry and spectroscopy,
only a single bolometric luminosity estimate by \citet{Currie:2010ju} exists in the literature:  
$\log(L/L_{\odot})$ = --6.1$^{+0.3}_{-0.2}$ dex.  
To improve on this value we use use existing near-infrared measurements to constrain the
1--3 $\mu$m SED together with atmospheric models for a bolometric correction.
We first anchor the 1.0--1.75 $\mu$m spectrum of Gl 758 B from \citet{Nilsson:2017hm}
by flux calibrating the P1640 observations to the $H$-band apparent magnitude from \citet{Thalmann:2009ca}.
To this we add the photometry from \citet{Vigan:2016gq} and \citet{Janson:2011dh}
to directly account for the comparably high $K$-band flux from this object.
A solar-metallicity BT-Settl ``CIFIST2011'' atmospheric model 
with $T_\mathrm{eff}$=650~K and $\log g$=5.0 dex is used
for the long-wavelength bolometric correction (2.5--500~$\mu$m) by flux-calibrating the
model to the $L'$-band photometry from \citet{Janson:2011dh}.
This approach also agrees with the $M_S$-band upper limit from \citet{Janson:2011dh}.
The same model is used for the short-wavelength correction (0.1--1.0 $\mu$m) by scaling
that region to the blue end of the P1640 spectrum.  
Uncertainties in the spectral measurements, photometry, and flux calibration scale factors 
for the model and spectrum are all
accounted for in a Monte Carlo fashion by integrating under new realizations of the complete
0.1--500~$\mu$m spectrum.
This procedure yields a bolometric luminosity of $\log(L/L_{\odot})$ = --6.07 $\pm$ 0.03 dex for Gl 758 B.
To assess possible systematic errors, we experimented with alternative atmospheric models 
from the same grid with effective temperatures of 600~K and 700~K.
The results following the same procedure are within 0.02 dex of the value we obtained 
with the 650 K model, which is
smaller than the impact of random measurement errors.

\subsection{Comparison with Evolutionary Models}{\label{sec:evmod}}

With a measured luminosity, age, and dynamical mass, Gl 758 B offers a rare opportunity to test 
substellar evolutionary models.
For this analysis we begin with the assumption that the age range spans 1--6~Gyr following \citet{Vigan:2016gq},
but ultimately re-evaluate this constraint based on recent results from isochrone fitting.
We select a variety of publicly available models  from the 
literature for this exercise: the Cond models
from \citet{Baraffe:2003bj}; three versions of
evolutionary models from \citet{Saumon:2008im}
with no clouds (``SM-NC''), a hybrid prescription for the evolution of clouds at the L/T transition (``SM-Hybrid''),
and the retention of thick clouds at all temperatures (``SM-f2''); and the grid from  \citet{Burrows:1997jq}.
All have solar compositions.
These models mainly differ in their treatment of atmospheric clouds and molecular opacities, which act
as boundary conditions that control the evolution of brown dwarfs and giant planets as these objects 
radiatively cool over time (see, e.g.,  \citealt{Burrows:2001wq} and \citealt{Marley:2015bj} for detailed reviews).

Our approach for comparing the models to the observations utilizes a one-tailed hypothesis test.
We adopt a null hypothesis in which the
posterior probability density function for the dynamical mass of Gl 758 B is statistically consistent with the 
inferred mass distribution from evolutionary models at some threshold probability; we choose 0.3\% 
(within 3 $\sigma$) for this study.  
In other words, we calculate the probability that random draws from
the dynamical mass distribution differ from the inferred model-based mass distribution.  If these two 
values disagree by at least 0.997, then the null hypothesis is rejected and the two distributions 
are considered to be inconsistent with each other.


\begin{figure}
  \vskip -.4 in
  \hskip -0.3 in
  \resizebox{4.in}{!}{\includegraphics{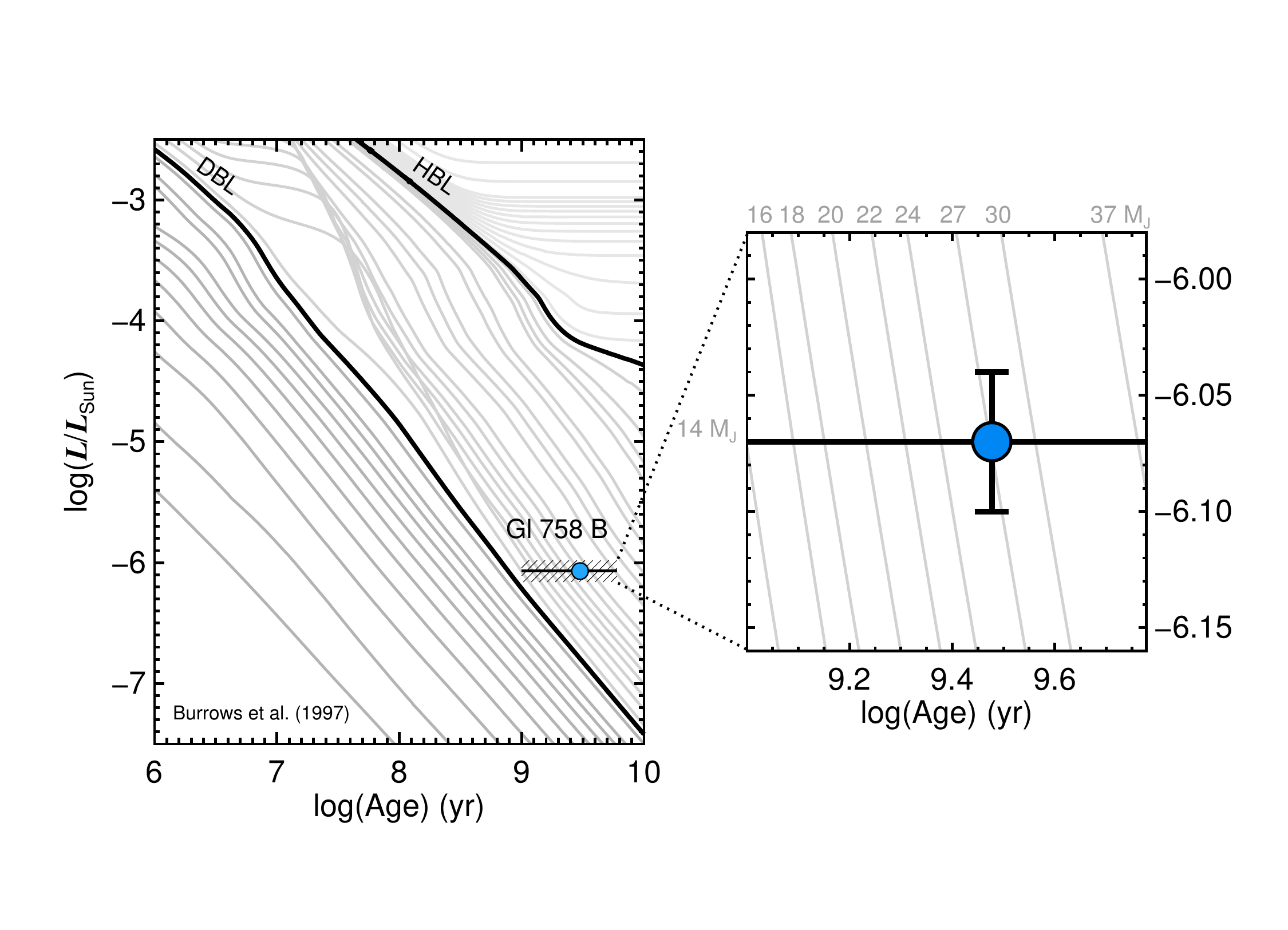}}
  \vskip -.4 in
  \caption{Luminosity and age of Gl 758 B with respect to evolutionary models from \citet{Burrows:1997jq}.
  The inset shows iso-mass tracks spanning the nominal age range of 1--6 Gyr for the host star.
  The hydrogen-burning limit (HBL) and deuterium-burning limit (DBL) are labeled.
    \label{fig:lumage} } 
\end{figure}

For each set of evolutionary models, we randomly draw ages from a uniform distribution
($\tau$ = $\mathcal{U}(1,6)$ Gyr) and luminosities from a log-normal distribution 
($\log (L/L_{\odot})$ = $\mathcal{N}(\mu$=$-6.07,\sigma$=$0.03)$ dex); a visual reference  
is shown in Figure~\ref{fig:lumage} for the Burrows models.  For each \{$\tau$,$L$\} pair we 
find the corresponding mass by finely interpolating the model grid.  This Monte Carlo process
is repeated of order 10$^6$ times to create a distribution of expected masses for 
each model.  The predicted and dynamical mass distributions are then quantitatively compared for consistency.

\begin{figure}
  \vskip -1.2 in
  \hskip -0.9 in
  \resizebox{5.1in}{!}{\includegraphics{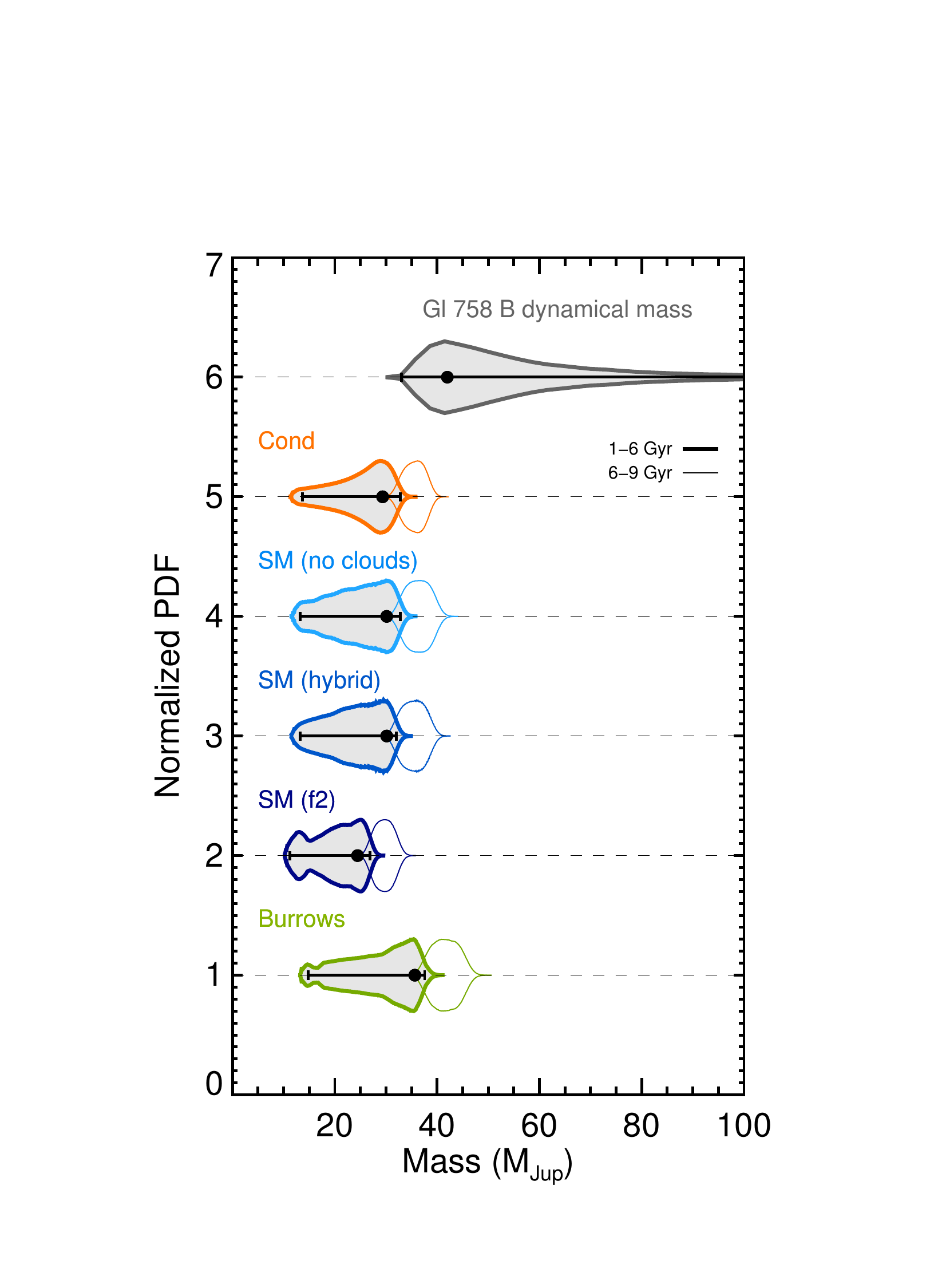}} 
  \vskip -.5 in
  \caption{Probability distributions for the inferred mass of Gl 758 B from
  five grids of evolutionary models compared to the dynamical mass from our orbit fit (top distribution).  
  The inferred mass distributions are calculated from the luminosity of Gl 758 B and 
  the nominal age range of 1--6~Gyr for the host star (thick lines).  Only the Burrows models
  formally agree with the dynamical mass at this age.  Somewhat older ages for the host star render the models in
  much better agreement with the observations (thin lines).   \label{fig:massprediction} } 
\end{figure}

Results from this analysis are listed in Table~\ref{tab:evmods} and illustrated in Figure~\ref{fig:massprediction}.
Based on the input age and luminosity distributions together with our threshold criterion for agreement,
only the Burrows models are formally consistent with the dynamical mass distribution--- which peaks
at 42~\Mjup \ and has a robust (4 $\sigma$) lower limit of 30.5~\Mjup.
However, even the formal agreement with the Burrows models is marginal and a slightly lower 
threshold for consistency would have rejected the null hypothesis; 
random draws from the dynamical mass distribution result in higher masses 99\% of the time.
All other models fail our hypothesis test.
The SM-f2 models disagree the most with the dynamical mass of Gl 758 B.  
By retaining clouds to temperatures well below the rainout limit for various grains, 
this prescription is the most unrealistic for T dwarfs like Gl 758 B (which is expected to be cloud-free)
so this result is unsurprising.


\begin{deluxetable*}{lccccccc}
\renewcommand\arraystretch{1.1}
\tabletypesize{\small}
\setlength{ \tabcolsep } {.1cm} 
\tablewidth{0pt}
\tablecolumns{8}
\tablecaption{Predictions from Evolutionary Models \label{tab:evmods}}
\tablehead{
       \multicolumn{3}{c}{Constraint} &  \multicolumn{5}{c}{Model} \\
     \hline
    \colhead{Age} &  \colhead{Luminosity}  & \colhead{Mass} &  &  & & &  \\
        \colhead{(Gyr)} &  \colhead{($\log(L/L_{\odot})$)}  & \colhead{(\Mjup)} &  \colhead{Cond}  & \colhead{SM-NC}  & \colhead{SM-Hybrid}  & \colhead{SM-f2}  & \colhead{Burrows}    
    }
\startdata
 & & &  \multicolumn{5}{c}{Predicted Mass (95.4\% Credible Interval)} \\
       \cline{4-8}
$\mathcal{U}$(1,6)    & $\mathcal{N}(-6.07,0.03)$  & $\cdots$  & 14--33 \Mjup & 13--33 \Mjup  & 13--32 \Mjup  & 11--27 \Mjup & 15--38 \Mjup \\
 & & &  \multicolumn{5}{c}{Minimum Compatible Age (99.7\% Lower Limit)} \\
       \cline{4-8}
$\cdots$    & $\mathcal{N}(-6.07,0.03)$  & PDF$({M_\mathrm{comp}})$  &  $>$6.2 Gyr &  $>$6.1 Gyr & $>$6.4 Gyr & $>$9.2 Gyr & $>$4.4 Gyr \\
 & & &  \multicolumn{5}{c}{Minimum Compatible Luminosity (99.7\% Lower Limit)} \\
       \cline{4-8}
$\mathcal{U}$(1,6)    & $\cdots$  & PDF$({M_\mathrm{comp}})$  & $>$--5.96 dex &  $>$--5.95 dex & $>$--5.92 dex & $>$--5.76 dex & $>$--6.13 dex \\
\enddata
\tablecomments{Predictions from evolutionary models for various permutations of input age, luminosity, and mass distributions.  Here $\mathcal{U}$($a$,$b$) refers to a linearly-uniform distribution from $a$ to $b$, $\mathcal{N}$($\mu$,$\sigma$) is a normal distribution with a mean $\mu$ and standard deviation $\sigma$, and PDF$({M_\mathrm{comp}})$ is our measured probability density function for the dynamical mass of Gl 758 B.}
\end{deluxetable*}

The general tension between the models and our dynamical mass measurement may 
point to physics not yet incorporated into current substellar models,  
which could originate from several sources: 
interior physics and thermal structure; sources of atmospheric opacity and their
evolution with temperature; initial entropy and accretion history of Gl 758 B; or ill-matched 
metallicities of the models and the companion.
On the other hand, it is also possible that the discrepancy originates from the observational side,
most likely with the age of the system.  For our default analysis we adopted the 1--6~Gyr estimate 
by \citet{Vigan:2016gq} based on isochrone fitting of the host star (which resulted in younger 
ages of $\approx$1--4~Gyr) and activity indicators (which resulted in older ages of $\approx$3--8 Gyr).
Older ages result in higher predicted masses for the same luminosity, so this could also be a natural explanation
for the disagreement.

To explore the possibility of an older age or a systematic offset in the luminosities of the models, 
we perform a series of tests to identify the minimum compatible ages and the minimum compatible luminosities
that render the inferred and dynamical mass distributions into agreement.
For the former, we begin by randomly drawing masses from interpolated evolutionary model grids 
following a normal distribution in log-luminosity ($\mathcal{N}(-6.07,0.03)$ dex) and 
fixing the  starting age at 1.0 Gyr.  Using the same threshold requirement of 0.3\%, we compare the inferred 
and dynamical mass distributions for consistency.  This process is then repeated by increasing the 
age in 0.1 Gyr steps until the mass distributions agree at the threshold level.  
Results are summarized in Table~\ref{tab:evmods}; the Cond, SM-NC, and SM-Hybrid
models all imply similar ages of $\gtrsim$6 Gyr.  The SM-f2 grid is only consistent with the dynamical
mass for extremely old ages of $\gtrsim$9~Gyr, and the Burrows models agree for ages beyond 4.4~Gyr.

A similar process is carried out to identify the minimum luminosity consistent with the measured mass.
We randomly draw masses based on a uniform distribution of ages ($\mathcal{U}(1,6)$ Gyr) and 
a fixed starting luminosity of --6.50~dex, then test the inferred and dynamical mass distributions for consistency.
The luminosity is increased in increments of 0.01 dex until agreement is reached.  
If the 1--6 Gyr age estimate is correct, that would imply that the Cond, SM-NC, SM-Hybrid, and SM-f2
evolutionary models are overluminous by 0.11 dex, 0.12 dex, 0.15 dex, and 0.31 dex, 
respectively (Table~\ref{tab:evmods}).
This potential discrepancy is in the opposite sense from results by \citet{Dupuy:2014iz} and \citet{Dupuy:2009jq},
who found that substellar cooling models under-predict the luminosities of brown dwarfs with dynamical masses 
by $\approx$0.2--0.4 dex, at least at relatively young age of $\approx$0.5--1~Gyr.

Altogether, the most likely culprit for the disagreement in mass probably resides in the age of Gl 758.  Older ages of 6--9 Gyr 
would readily put the predicted and dynamical distributions in excellent agreement and are indeed
suggested from the low activity level, lack of X-ray emission, enhanced metallicity, and slow projected rotational velocity of Gl 758 
(\citealt{Mamajek:2008jz}; \citealt{Thalmann:2009ca}; \citealt{Vigan:2016gq}).
Although there are a wide range of age estimates for the host star from isochrone fitting in the literature, 
more recent analyses are converging on an older value that agrees better with activity indicators. 
For example, a recent study by \citet{Luck:2016jd} found an average age of 7.5 Gyr (with a range of 5.3 Gyr about that value) 
using four sets of isochrones, and
\citet{Brewer:2016gf} find an isochronal age of 7.5 Gyr with a range of 4.6--10.4 Gyr using the Yonsei-Yale models.

\subsection{Limits on Planetary Companions}{\label{sec:rvresiduals}}


\begin{figure}
  \vskip -.1 in
  \hskip -.3 in
  \resizebox{4.2in}{!}{\includegraphics{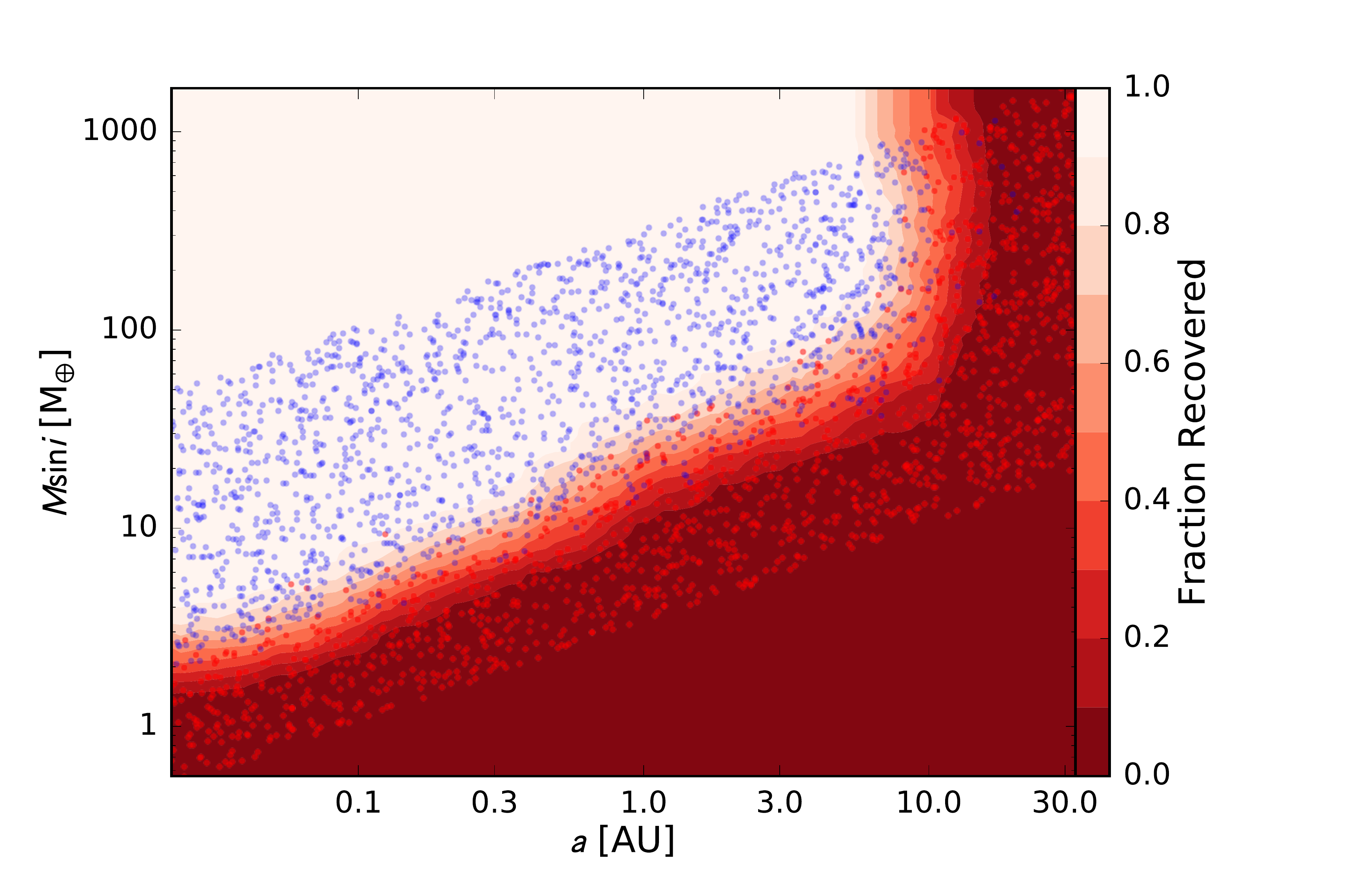}} 
  \vskip -.1 in
  \caption{ Sensitivity map for close-in planets orbiting Gl 758 based on residual
  RVs after removing the best-fit orbit for Gl 758 B.  Blue dots represent injected companions
  that were recovered following the procedure described in \citet{Howard:2016gc}.  Red dots represent injected planets
  that were not recovered, and contours delineate the fraction of injected planets that were recovered.    \label{fig:limits} } 
\end{figure}

We searched the residual RVs for closer-in planet candidates using a Lomb-Scargle periodogram
after removing the best-fit orbital solution of Gl 758 B.
The strongest power from 1--10$^4$ days is at a period
of 245.5 days, but that potential signal has a false alarm probability of 0.4\% and its corresponding
velocity semi-amplitude is at the level of the noise in the data, so it is unlikely to be real.  
No frequencies have powers that exceed our threshold false alarm probability of 0.1\% for planet detection. 
We conclude that there is no convincing evidence of any close-in planet candidates in our data.

Detection limits are quantified using injection-recovery tests as described in \citet{Howard:2016gc}.
Synthetic planets on circular orbits are sequentially injected into the RV residuals by randomly drawing 
pairs of minimum mass and period surrounding the detection threshold.  
A periodogram is used to search for planets within each artificial dataset with a 1\% false alarm probability 
threshold for recovery and the requirement of a similar period and phase as the injected planet.
Results are shown in Figure~\ref{fig:limits}.
Gl 758 is devoid of close-in giant planets ($\gtrsim$100~\Mearth) within 10~AU, 
sub-Saturns ($\approx$10--100~\Mearth) within 3~AU,
and super-Earths ($\approx$2--10~\Mearth) interior to 0.1~AU.

Given its periastron passage of 8.9$^{+3.7}_{-2.0}$ AU, 
it is likely that Gl 758 B has impacted the formation efficiency and dynamical stability of closer-in planets in this system.
If Gl 758 B formed relatively quickly ($\lesssim$1~Myr), perhaps from turbulent fragmentation (e.g., \citealt{Bate:2009br}) 
with subsequent migration to its current orbit, 
this implies that the circumprimary protoplanetary disk would have been truncated between $\sim$0.2--0.35$a$, or about 4--7 AU (\citealt{Artymowicz:1994bw}). 
The lack of planets outside of this region is not unexpected, but planet formation interior to this region may still have been possible
(e.g., Kepler-444; \citealt{Campante:2015ei}; \citealt{Dupuy:2016dh}).

However, the lack of planets at small orbital distances from Gl 758 is not particularly unusual compared
to the statistical properties of planets orbiting GK dwarfs in general.
For example, \citet{Cumming:2008hg} find that about 10\% of Sun-like stars host
giant planets with minimum masses between 0.3--10~\Mjup \ within $\approx$3 AU.
This value increases to about 14\% by extrapolating the planet period distribution out to 10 AU.
\citet{Wittenmyer:2016hp} infer the frequency of Jupiter analogs between 3--7~AU around 
solar-type stars to be 6.2$^{+2.8}_{-1.6}$\% (see also \citealt{Wittenmyer:2011fw} and \citealt{Zechmeister:2013gt}).
\citet{Petigura:2013ij} measure a completeness-corrected frequency of about 55\% for all planets orbiting
GK stars between 1--12~$R_{\earth}$ and orbital periods from 5-100 days (0.05--0.4~AU),
with lower-mass planets outnumbering gas giants by a factor of $\approx$50:1
(see also \citealt{Howard:2012di}; \citealt{Fressin:2013df}; \citealt{Youdin:2011gz}).
While it is possible that the absence of close-in planets could be related to the presence of Gl 758 B,
this apparent desert is also broadly consistent with the 
overall statistical properties of single FGK stars.


\begin{figure}
  \vskip -.8 in
  \hskip -1.2 in
  \resizebox{5.6in}{!}{\includegraphics{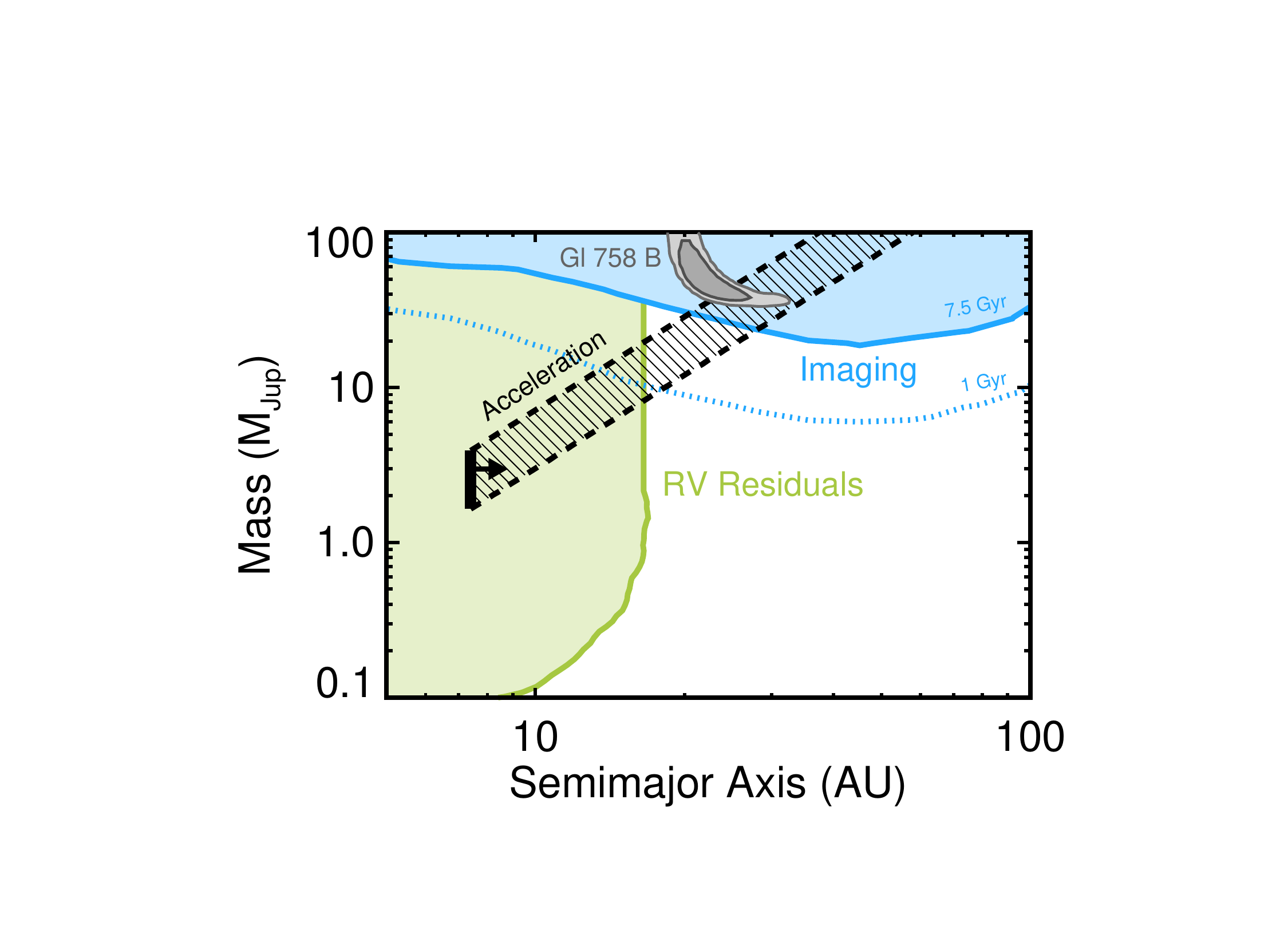}} 
  \vskip -.5 in
  \caption{Overview of  constraints on the Gl 758 system.  The shaded blue and green regions 
  illustrate the masses and separations over which we are sensitive to companions at the 10\% level based on
  our imaging observations and RV residuals after subtracting out the secular acceleration.
  The hashed area denotes the 95\% interval of minimum masses and separations consistent with the measured long-baseline
  acceleration of 2--5~m s yr$^{-1}$.  The cutoff at 7.5~AU  corresponds to the minimum period of the perturbing 
  companion, limited by the baseline of our McDonald RVs. 
  Our joint constraint on the dynamical mass and semimajor axis for Gl 758 B is shown in gray, which assumes 
  the acceleration originates entirely from this object.    
  Note that the imaging ``exclusion zone'' assumes hot-start evolutionary models from \citet{Baraffe:2003bj}, an age of 7.5~$\pm$ 1.5~Gyr
  for the system, and circular orbits.  The corresponding contrast curve is flux calibrated using the host star and coronagraph 
  throughput attenuation from \citet{Bowler:2015ja}. For comparison, the dotted blue line corresponds to the 10\% sensitivity contour for a 
  younger age of 1 Gyr.
\label{fig:exclusion} } 
\end{figure}

Figure~\ref{fig:exclusion} shows the mass and separation regimes over which our imaging data and
the residuals from our RVs are sensitive.  
Together this rules out giant planet and brown dwarf companions at close separations as well as massive companions at wide
orbital distances.  However, there exists a large region beyond about 10 AU and at masses less than about 30~\Mjup \ 
where additional companions could evade detection, assuming an older age of about 6--9~Gyr.  
If a giant planet or another brown dwarf resides in this system---which remains possible both below 
our detection limits or simply at unfavorable viewing geometries---the acceleration 
we observe would be the superposition from one or more additional companions besides Gl 758 B.
This could potentially influence the amplitude and shape of the evolving acceleration and may even
reconcile the dynamical mass measurement and the younger age. 
At this point there are no signs of another companion, but continued RV monitoring
and deeper high-contrast imaging would be beneficial to further map the architecture of this system.

\section{Summary and Conclusion}{\label{sec:conclusions}}

We have used 630 RV observations of the G8 star Gl~758 taken over the past two decades 
together with seven years of
astrometry with NIRC2 to measure the dynamical mass of the T7--T8 brown dwarf companion Gl~758~B.
A joint fit of the RVs and our new astrometry with a 15-parameter Keplerian model
yields a mass of 42$^{+19}_{-7}$~\Mjup \ for the companion, assuming a host star prior of 0.97~$\pm$0.02 \Msun, 
with a strict lower limit of 30.5~\Mjup \ 
and a long tail to higher masses.  Gl 758 B orbits its host about once a century with a modest eccentricity
between about 0.26--0.67 and a semimajor axis of 21~AU.
Based on our best-fit orbit solution, we expect the host star's acceleration to continue to steepen over the next
several years and then reverse sign in a few decades. 
Despite the excellent RV precision and long observational baseline, no close-in planets are detected
in the RV residuals.
Based on a revised bolometric luminosity for Gl 758 B, 
most evolutionary models are inconsistent with the companion's dynamical mass for ages less than 6~Gyr.

Continued ground-based RV and astrometric orbit monitoring will progressively improve the orbit and mass constraint 
of Gl 758 B.  In addition, the astrometric perturbation of Gl 758~B on its host star 
should be readily detected by \emph{Gaia} during its nominal 5-year mission lifetime.
This will dramatically refine the inclination and dynamical mass of Gl 758 B.
Similarly, more precise age determinations for the host star will enable more robust tests of
evolutionary models approaching the planetary-mass regime.

Gl 758 B is the lowest-mass companion inducing a measured acceleration on its host star,
demonstrating the continued value and productivity of long-baseline RV planet searches.
In the future, the combination of RVs and direct imaging will regularly yield 
dynamical masses for exoplanets using the next generation of 
ground-based 30-m class telescopes and space-based missions like \emph{JWST} and \emph{WFIRST}.

\acknowledgements

We are especially grateful to the many dedicated observers and support staff
at McDonald Observatory, Keck, and APF who contributed
to this expansive dataset over the years, in particular
S. Barnes,
I. Ramirez,
E. Brugamyer,
C. Caldwell,
K. Gullikson,
P. Robertson, 
G. Marcy, 
H. Isaacson, 
D. Fischer, 
K. Peek, 
E. Sinukoff, 
J. Johnson, 
L. Weiss, 
G. Torres, 
G. Bakos, 
T. Morton, 
J. Brewer, 
S. Pineda, 
J. Wang, 
C. Clanton, 
M. Bottom, 
and J. Curtis.  
This study would not have been possible without this communal assistance and resourcefulness.
We thank Chris Sneden for many helpful discussions on the properties of the host star and
Ricky Nilsson for sharing the spectrum of Gl 758 B from P1640.
The McDonald Observatory Planet Search (ME, WDC, PJM) is supported by the 
National Science Foundation through grant AST-1313075.
This work was supported by a NASA Keck PI Data Award, administered by the NASA Exoplanet Science Institute. Data presented herein were obtained at the W. M. Keck Observatory from telescope time allocated to the National Aeronautics and Space Administration through the agency's scientific partnership with the California Institute of Technology and the University of California. 
The Observatory was made possible by the generous financial support of the W. M. Keck Foundation.
Support for this work was provided by NASA through Hubble Fellowship grants HST-HF2-51369.001-A 
awarded by the Space Telescope Science Institute, which is operated by the Association of Universities for Research in Astronomy, Inc., for NASA, under contract NAS5-26555. 
This paper includes data taken at The McDonald Observatory of The University of Texas at Austin.
The authors wish to recognize and acknowledge the very significant cultural role and reverence that the summit of Mauna Kea has always had within the indigenous Hawaiian community.  We are most fortunate to have the opportunity to conduct observations from this mountain.

\facility{Keck:II (NIRC2), Keck:I (HIRES), Smith (Tull Coud\'{e} Spectrograph), APF (Levy)}



\begin{thebibliography}{}
\expandafter\ifx\csname natexlab\endcsname\relax\def\natexlab#1{#1}\fi

\bibitem[{Artymowicz \& Lubow(1994)}]{Artymowicz:1994bw}
Artymowicz, P., \& Lubow, S.~H. 1994, Astrophys. J., 421, 651

\bibitem[{Baraffe {et~al.}(2003)Baraffe, Chabrier, Barman, Allard, \&
  Hauschildt}]{Baraffe:2003bj}
Baraffe, I., Chabrier, G., Barman, T.~S., Allard, F., \& Hauschildt, P.~H.
  2003, A{\&}A, 402, 701

\bibitem[{Bate(2009)}]{Bate:2009br}
Bate, M.~R. 2009, Monthly Notices RAS, 392, 590

\bibitem[{Bowler(2016)}]{Bowler:2016jk}
Bowler, B.~P. 2016, Publications of the Astronomical Society of the Pacific,
  128, 102001

\bibitem[{Bowler {et~al.}(2015{\natexlab{a}})Bowler, Liu, Shkolnik, \&
  Tamura}]{Bowler:2015ja}
Bowler, B.~P., Liu, M.~C., Shkolnik, E.~L., \& Tamura, M. 2015{\natexlab{a}},
  The Astrophysical Journal Supplement Series, 216, 7

\bibitem[{Bowler {et~al.}(2015{\natexlab{b}})Bowler, Shkolnik, Liu, Schlieder,
  Mann, Dupuy, Hinkley, Crepp, Johnson, Howard, Flagg, Weinberger, Aller,
  Allers, Best, Kotson, Montet, Herczeg, Baranec, Riddle, Law, Nielsen, Wahhaj,
  Biller, \& Hayward}]{Bowler:2015ch}
Bowler, B.~P., Shkolnik, E.~L., Liu, M.~C., {et~al.} 2015{\natexlab{b}}, The
  Astrophysical Journal, 806, 62

\bibitem[{Bowler {et~al.}(2017)Bowler, Liu, Mawet, Ngo, Malo, Mace, McLane, Lu,
  Tristan, Hinkley, Hillenbrand, Shkolnik, Benneke, \& Best}]{Bowler:2017hq}
Bowler, B.~P., Liu, M.~C., Mawet, D., {et~al.} 2017, The Astronomical Journal,
  153, 1

\bibitem[{Brewer {et~al.}(2016)Brewer, Fischer, Valenti, \&
  Piskunov}]{Brewer:2016gf}
Brewer, J.~M., Fischer, D.~A., Valenti, J.~A., \& Piskunov, N. 2016, The
  Astrophysical Journal Supplement Series, 225, 1

\bibitem[{Burrows {et~al.}(2001)Burrows, Hubbard, Lunine, \&
  Liebert}]{Burrows:2001wq}
Burrows, A., Hubbard, W.~B., Lunine, J.~I., \& Liebert, J. 2001, Reviews of
  Modern Physics, 73, 719

\bibitem[{Burrows {et~al.}(1997)Burrows, Marley, Hubbard, Lunine, Guillot,
  Saumon, Freedman, Sudarsky, \& Sharp}]{Burrows:1997jq}
Burrows, A., Marley, M., Hubbard, W.~B., {et~al.} 1997, Astrophysical Journal,
  491, 856

\bibitem[{Campante {et~al.}(2015)Campante, Barclay, Swift, Huber, Adibekyan,
  Cochran, Burke, Isaacson, Quintana, Davies, Silva~Aguirre, Ragozzine, Riddle,
  Baranec, Basu, Chaplin, Christensen-Dalsgaard, Metcalfe, Bedding, Handberg,
  Stello, Brewer, Hekker, Karoff, Kolbl, Law, Lundkvist, Miglio, Rowe, Santos,
  Van~Laerhoven, Arentoft, Elsworth, Fischer, Kawaler, Kjeldsen, Lund, Marcy,
  Sousa, Sozzetti, \& White}]{Campante:2015ei}
Campante, T.~L., Barclay, T., Swift, J.~J., {et~al.} 2015, ApJ, 799, 170

\bibitem[{Cheetham {et~al.}(2017)Cheetham, S{\'e}gransan, Peretti, Delisle,
  Hagelberg, Beuzit, Forveille, Marmier, Udry, \& Wildi}]{Cheetham:2017wb}
Cheetham, A., S{\'e}gransan, D., Peretti, S., {et~al.} 2017, arXiv,
  arXiv:1712.05217

\bibitem[{Cochran {et~al.}(1997)Cochran, Hatzes, \& Butler}]{Cochran:1997ta}
Cochran, W.~D., Hatzes, A.~P., \& Butler, R.~P. 1997, Astrophysical Journal,
  483, 457

\bibitem[{Crepp {et~al.}(2016)Crepp, Gonzales, Bechter, Montet, Johnson,
  Piskorz, Howard, \& Isaacson}]{Crepp:2016fg}
Crepp, J.~R., Gonzales, E.~J., Bechter, E.~B., {et~al.} 2016, ApJ, 831, 1

\bibitem[{Crepp {et~al.}(2014)Crepp, Johnson, Howard, Marcy, Brewer, Fischer,
  Wright, \& Isaacson}]{Crepp:2014ce}
Crepp, J.~R., Johnson, J.~A., Howard, A.~W., {et~al.} 2014, ApJ, 781, 29

\bibitem[{Crepp {et~al.}(2012)Crepp, Johnson, Fischer, Howard, Marcy, Wright,
  Isaacson, Boyajian, Von~Braun, Hillenbrand, Hinkley, Carpenter, \&
  Brewer}]{Crepp:2012eg}
Crepp, J.~R., Johnson, J.~A., Fischer, D.~A., {et~al.} 2012, The Astrophysical
  Journal, 751, 97

\bibitem[{Cumming {et~al.}(2008)Cumming, Butler, Marcy, Vogt, Wright, \&
  Fischer}]{Cumming:2008hg}
Cumming, A., Butler, R.~P., Marcy, G.~W., {et~al.} 2008, PASP, 120, 531

\bibitem[{Currie {et~al.}(2010)Currie, Bailey, Fabrycky, Murray-Clay, Rodigas,
  \& Hinz}]{Currie:2010ju}
Currie, T., Bailey, V., Fabrycky, D., {et~al.} 2010, The Astrophysical Journal,
  721, L177

\bibitem[{Dupuy {et~al.}(2016)Dupuy, Kratter, Kraus, Isaacson, Mann, Ireland,
  Howard, \& Huber}]{Dupuy:2016dh}
Dupuy, T.~J., Kratter, K.~M., Kraus, A.~L., {et~al.} 2016, ApJ, 817, 1

\bibitem[{Dupuy \& Liu(2017)}]{Dupuy:2017ke}
Dupuy, T.~J., \& Liu, M.~C. 2017, The Astrophysical Journal Supplement Series,
  231, 0

\bibitem[{Dupuy {et~al.}(2009)Dupuy, Liu, \& Ireland}]{Dupuy:2009jq}
Dupuy, T.~J., Liu, M.~C., \& Ireland, M.~J. 2009, The Astrophysical Journal,
  692, 729

\bibitem[{Dupuy {et~al.}(2014)Dupuy, Liu, \& Ireland}]{Dupuy:2014iz}
---. 2014, ApJ, 790, 133

\bibitem[{Earl \& Deem(2005)}]{Earl:2005hv}
Earl, D.~J., \& Deem, M.~W. 2005, Phys. Chem. Chem. Phys., 7, 3910

\bibitem[{Endl {et~al.}(2000)Endl, K{\"u}rster, \& Els}]{Endl:2000ui}
Endl, M., K{\"u}rster, M., \& Els, S. 2000, A{\&}A, 362, 585

\bibitem[{Endl {et~al.}(2016)Endl, Brugamyer, Cochran, MacQueen, Robertson,
  Meschiari, Ram{\'\i}rez, Shetrone, Gullikson, Johnson, Wittenmyer, Horner,
  Ciardi, Horch, Simon, Howell, Everett, Caldwell, \&
  Castanheira}]{Endl:2016kk}
Endl, M., Brugamyer, E.~J., Cochran, W.~D., {et~al.} 2016, ApJ, 818, 1

\bibitem[{Foreman-Mackey {et~al.}(2013)Foreman-Mackey, Hogg, Lang, \&
  Goodman}]{ForemanMackey:2013io}
Foreman-Mackey, D., Hogg, D.~W., Lang, D., \& Goodman, J. 2013, Publications of
  the Astronomical Society of the Pacific, 125, 306

\bibitem[{Fressin {et~al.}(2013)Fressin, Torres, Charbonneau, Bryson,
  Christiansen, Dressing, Jenkins, Walkowicz, \& Batalha}]{Fressin:2013df}
Fressin, F., Torres, G., Charbonneau, D., {et~al.} 2013, ApJ, 766, 81

\bibitem[{Fulton {et~al.}(2015)Fulton, Weiss, Sinukoff, Isaacson, Howard,
  Marcy, Henry, Holden, \& Kibrick}]{Fulton:2015gj}
Fulton, B.~J., Weiss, L.~M., Sinukoff, E., {et~al.} 2015, ApJ, 805, 1

\bibitem[{{Gaia Collaboration} {et~al.}(2016){Gaia Collaboration}, Brown,
  Vallenari, Prusti, de~Bruijne, Mignard, Drimmel, Babusiaux, Bailer-Jones,
  Bastian, Biermann, Evans, Eyer, Jansen, Jordi, Katz, Klioner, Lammers,
  Lindegren, Luri, O~Mullane, Panem, Pourbaix, Randich, Sartoretti, Siddiqui,
  Soubiran, Valette, van Leeuwen, Walton, Aerts, Arenou, Cropper, H~g,
  Lattanzi, Grebel, Holland, Huc, Passot, Perryman, Bramante, Cacciari,
  Casta~eda, Chaoul, Cheek, De~Angeli, Fabricius, Guerra, Hern~ndez,
  Jean-Antoine-Piccolo, Masana, Messineo, Mowlavi, Nienartowicz, Ord~ez Blanco,
  Panuzzo, Portell, Richards, Riello, Seabroke, Tanga, Th~venin, Torra, Els,
  Gracia-Abril, Comoretto, Garcia-Reinaldos, Lock, Mercier, Altmann, Andrae,
  Astraatmadja, Bellas-Velidis, Benson, Berthier, Blomme, Busso, Carry,
  Cellino, Clementini, Cowell, Creevey, Cuypers, Davidson, De~Ridder,
  de~Torres, Delchambre, Dell~Oro, Ducourant, Fr~mat, Garc~a Torres, Gosset,
  Halbwachs, Hambly, Harrison, Hauser, Hestroffer, Hodgkin, Huckle, Hutton,
  Jasniewicz, Jordan, Kontizas, Korn, Lanzafame, Manteiga, Moitinho, Muinonen,
  Osinde, Pancino, Pauwels, Petit, Recio-Blanco, Robin, Sarro, Siopis, Smith,
  Smith, Sozzetti, Thuillot, van Reeven, Viala, Abbas, Abreu~Aramburu, Accart,
  Aguado, Allan, Allasia, Altavilla, lvarez, Alves, Anderson, Andrei,
  Anglada~Varela, Antiche, Antoja, Ant~n, Arcay, Bach, Baker, Balaguer-N~ez,
  Barache, Barata, Barbier, Barblan, Barrado~y Navascu~s, Barros, Barstow,
  Becciani, Bellazzini, Bello Garc~a, Belokurov, Bendjoya, Berihuete, Bianchi,
  Bienaym, Billebaud, Blagorodnova, Blanco-Cuaresma, Boch, Bombrun, Borrachero,
  Bouquillon, Bourda, Bouy, Bragaglia, Breddels, Brouillet, Br~semeister,
  Bucciarelli, Burgess, Burgon, Burlacu, Busonero, Buzzi, Caffau, Cambras,
  Campbell, Cancelliere, Cantat-Gaudin, Carlucci, Carrasco, Castellani,
  Charlot, Charnas, Chiavassa, Clotet, Cocozza, Collins, Costigan, Crifo,
  Cross, Crosta, Crowley, Dafonte, Damerdji, Dapergolas, David, David, De~Cat,
  de~Felice, de~Laverny, De~Luise, De~March, de~Martino, de~Souza, Debosscher,
  del Pozo, Delbo, Delgado, Delgado, Di~Matteo, Diakite, Distefano, Dolding,
  Dos~Anjos, Drazinos, Duran, Dzigan, Edvardsson, Enke, Evans, Eynard~Bontemps,
  Fabre, Fabrizio, Faigler, Falc~o, Farr~s Casas, Federici, Fedorets, Fern~ndez
  Hern~ndez, Fernique, Fienga, Figueras, Filippi, Findeisen, Fonti, Fouesneau,
  Fraile, Fraser, Fuchs, Gai, Galleti, Galluccio, Garabato, Garc~a Sedano,
  Garofalo, Garralda, Gavras, Gerssen, Geyer, Gilmore, Girona, Giuffrida,
  Gomes, Gonz~lez Marcos, Gonz~lez N~ez, Gonz~lez Vidal, Granvik, Guerrier,
  Guillout, Guiraud, G~rpide, Guti~rrez S~nchez, Guy, Haigron, Hatzidimitriou,
  Haywood, Heiter, Helmi, Hobbs, Hofmann, Holl, Holland, Hunt, Hypki, Icardi,
  Irwin, Jevardat~de Fombelle, Jofr, Jonker, Jorissen, Julbe, Karampelas,
  Kochoska, Kohley, Kolenberg, Kontizas, Koposov, Kordopatis, Koubsky,
  Krone-Martins, Kudryashova, Kull, Bachchan, \&
  Lacos...}]{GaiaCollaboration:2016gd}
{Gaia Collaboration}, Brown, A. G.~A., Vallenari, A., {et~al.} 2016, A{\&}A,
  595, A2

\bibitem[{Howard \& Fulton(2016)}]{Howard:2016gc}
Howard, A.~W., \& Fulton, B.~J. 2016, Publications of the Astronomical Society
  of the Pacific, 128, 1

\bibitem[{Howard {et~al.}(2010)Howard, Johnson, Marcy, Fischer, Wright, Bernat,
  Henry, Peek, Isaacson, Apps, Endl, Cochran, Valenti, Anderson, \&
  Piskunov}]{Howard:2010dia}
Howard, A.~W., Johnson, J.~A., Marcy, G.~W., {et~al.} 2010, The Astrophysical
  Journal, 721, 1467

\bibitem[{Howard {et~al.}(2012)Howard, Marcy, Bryson, Jenkins, Rowe, Batalha,
  Borucki, Koch, Dunham, Gautier, Van~Cleve, Cochran, Latham, Lissauer, Torres,
  Brown, Gilliland, Buchhave, Caldwell, Christensen-Dalsgaard, Ciardi, Fressin,
  Haas, Howell, Kjeldsen, Seager, Rogers, Sasselov, Steffen, Basri,
  Charbonneau, Christiansen, Clarke, Dupree, Fabrycky, Fischer, Ford, Fortney,
  Tarter, Girouard, Holman, Johnson, Klaus, Machalek, Moorhead, Morehead,
  Ragozzine, Tenenbaum, Twicken, Quinn, Isaacson, Shporer, Lucas, Walkowicz,
  Welsh, Boss, Devore, Gould, Smith, Morris, Prsa, Morton, Still, Thompson,
  Mullally, Endl, \& MacQueen}]{Howard:2012di}
Howard, A.~W., Marcy, G.~W., Bryson, S.~T., {et~al.} 2012, The Astrophysical
  Journal Supplement Series, 201, 15

\bibitem[{Ireland {et~al.}(2008)Ireland, Kraus, Martinache, Lloyd, \&
  Tuthill}]{Ireland:2008kr}
Ireland, M.~J., Kraus, A., Martinache, F., Lloyd, J.~P., \& Tuthill, P.~G.
  2008, The Astrophysical Journal, 678, 463

\bibitem[{Janson {et~al.}(2011)Janson, Carson, Thalmann, McElwain, Goto, Crepp,
  Wisniewski, Abe, Brandner, Burrows, Egner, Feldt, Grady, Golota, Guyon,
  Hashimoto, Hayano, Hayashi, Hayashi, Henning, Hodapp, Ishii, Iye, Kandori,
  Knapp, Kudo, Kusakabe, Kuzuhara, Matsuo, Mayama, Miyama, Morino, Moro-Martin,
  Nishimura, Pyo, Serabyn, Suto, Suzuki, Takami, Takato, Terada, Tofflemire,
  Tomono, Turner, Watanabe, Yamada, Takami, Usuda, \& Tamura}]{Janson:2011dh}
Janson, M., Carson, J., Thalmann, C., {et~al.} 2011, ApJ, 728, 85

\bibitem[{Konopacky {et~al.}(2010)Konopacky, Ghez, Barman, Rice, Bailey, White,
  Mclean, \& Duch{\^e}ne}]{Konopacky:2010kr}
Konopacky, Q.~M., Ghez, A.~M., Barman, T.~S., {et~al.} 2010, The Astrophysical
  Journal, 711, 1087

\bibitem[{Lafreni{\`e}re {et~al.}(2007)Lafreni{\`e}re, Marois, Doyon, Nadeau,
  \& Artigau}]{Lafreniere:2007bg}
Lafreni{\`e}re, D., Marois, C., Doyon, R., Nadeau, D., \& Artigau, {\'E}. 2007,
  The Astrophysical Journal, 660, 770

\bibitem[{Liu {et~al.}(2008)Liu, Dupuy, \& Ireland}]{Liu:2008ib}
Liu, M.~C., Dupuy, T.~J., \& Ireland, M.~J. 2008, The Astrophysical Journal,
  689, 436

\bibitem[{Liu {et~al.}(2002)Liu, Fischer, Graham, Lloyd, Marcy, \&
  Butler}]{Liu:2002fx}
Liu, M.~C., Fischer, D.~A., Graham, J.~R., {et~al.} 2002, The Astrophysical
  Journal, 571, 519

\bibitem[{Luck(2017)}]{Luck:2016jd}
Luck, R.~E. 2017, The Astronomical Journal, 153, 1

\bibitem[{Lucy(2014)}]{Lucy:2014kr}
Lucy, L.~B. 2014, A{\&}A, 563, A126

\bibitem[{Mamajek \& Hillenbrand(2008)}]{Mamajek:2008jz}
Mamajek, E.~E., \& Hillenbrand, L.~A. 2008, The Astrophysical Journal, 687,
  1264

\bibitem[{Marcy \& Butler(1992)}]{Marcy:1992ix}
Marcy, G.~W., \& Butler, R.~P. 1992, PASP, 104, 270

\bibitem[{Marley \& Robinson(2015)}]{Marley:2015bj}
Marley, M.~S., \& Robinson, T.~D. 2015, Annu. Rev. Astro. Astrophys., 53, 279

\bibitem[{Marois {et~al.}(2006)Marois, Lafreni{\`e}re, Doyon, Macintosh, \&
  Nadeau}]{Marois:2006df}
Marois, C., Lafreni{\`e}re, D., Doyon, R., Macintosh, B., \& Nadeau, D. 2006,
  The Astrophysical Journal, 641, 556

\bibitem[{Marois {et~al.}(2010)Marois, Macintosh, \& V{\'e}ran}]{Marois:2010hs}
Marois, C., Macintosh, B., \& V{\'e}ran, J.-P. 2010, Proc. SPIE, 7736, 77361J

\bibitem[{McCaughrean {et~al.}(2004)McCaughrean, Close, Scholz, Lenzen, Biller,
  Brandner, Hartung, \& Lodieu}]{Mccaughrean:2004ey}
McCaughrean, M.~J., Close, L.~M., Scholz, R.-D., {et~al.} 2004, A{\&}A, 413,
  1029

\bibitem[{Nelder \& Mead(1965)}]{Nelder:1965tk}
Nelder, J.~A., \& Mead, R. 1965, The Computer Journal, 7, 308

\bibitem[{Nilsson {et~al.}(2017)Nilsson, Veicht, Godfrey, Rice, Aguilar, Pueyo,
  Roberts, Oppenheimer, Brenner, Luszcz-Cook, Bacchus, Beichman, Burruss, Cady,
  Dekany, Fergus, Hillenbrand, hinkley, King, Lockhart, Parry,
  Sivaramakrishnan, Soummer, Vasisht, Zhai, \& Zimmerman}]{Nilsson:2017hm}
Nilsson, R., Veicht, A., Godfrey, P. A.~G., {et~al.} 2017, ApJ, 838, 0

\bibitem[{Petigura \& Howard(2013)}]{Petigura:2013ij}
Petigura, E.~A., \& Howard, A.~W. 2013, in Proceedings of the {\ldots}

\bibitem[{Potter {et~al.}(2002)Potter, Mart{\'\i}n, Cushing, Baudoz, Brandner,
  Guyon, \& Neuh{\"a}user}]{Potter:2002ie}
Potter, D., Mart{\'\i}n, E.~L., Cushing, M.~C., {et~al.} 2002, The
  Astrophysical Journal, 567, L133

\bibitem[{Rajan {et~al.}(2017)Rajan, Rameau, De~Rosa, Marley, Graham,
  Macintosh, Marois, Morley, Patience, Pueyo, Saumon, Ward-Duong, Ammons,
  Arriaga, Bailey, Barman, Bulger, Burrows, Chilcote, Cotten, Czekala, Doyon,
  Duchene, Esposito, Fitzgerald, Follette, Fortney, Goodsell, Greenbaum, Hibon,
  Hung, Ingraham, Johnson-Groh, Kalas, Konopacky, LaFreniere, Larkin, Maire,
  Marchis, Metchev, Millar-Blanchaer, Morzinski, Nielsen, Oppenheimer, Palmer,
  Patel, Perrin, Poyneer, Rantakyr{\"o}, Ruffio, Savransky, Schneider,
  Sivaramakrishnan, Song, Soummer, Thomas, Vasisht, Wallace, Wang, Wiktorowicz,
  \& Wolff}]{Rajan:2017hq}
Rajan, A., Rameau, J., De~Rosa, R.~J., {et~al.} 2017, The Astronomical Journal,
  154, 0

\bibitem[{Saumon \& Marley(2008)}]{Saumon:2008im}
Saumon, D., \& Marley, M.~S. 2008, The Astrophysical Journal, 689, 1327

\bibitem[{Service {et~al.}(2016)Service, Lu, Campbell, Sitarski, Ghez, \&
  Anderson}]{Service:2016gk}
Service, M., Lu, J.~R., Campbell, R., {et~al.} 2016, Publications of the
  Astronomical Society of the Pacific, 128, 1

\bibitem[{Tamura(2016)}]{Tamura:2016jg}
Tamura, M. 2016, Proceedings of the Japan Academy. Ser. B: Physical and
  Biological Sciences, 92, 45

\bibitem[{Thalmann {et~al.}(2009)Thalmann, Carson, Janson, Goto, Mcelwain,
  Egner, Feldt, Hashimoto, Hayano, Henning, Hodapp, Kandori, Klahr, Kudo,
  Kusakabe, Mordasini, Morino, Suto, Suzuki, \& Tamura}]{Thalmann:2009ca}
Thalmann, C., Carson, J., Janson, M., {et~al.} 2009, The Astrophysical Journal,
  707, L123

\bibitem[{Torres(1999)}]{Torres:1999gc}
Torres, G. 1999, The Publications of the Astronomical Society of the Pacific,
  111, 169

\bibitem[{Tull {et~al.}(1995)Tull, MacQueen, \& Sneden}]{Tull:1995tn}
Tull, R.~G., MacQueen, P.~J., \& Sneden, C. 1995, Publications of the {\ldots}

\bibitem[{Valenti {et~al.}(1995)Valenti, Butler, \& Marcy}]{Valenti:1995bk}
Valenti, J.~A., Butler, R.~P., \& Marcy, G.~W. 1995, Publications of the
  Astronomical Society of the Pacific, 107, 966

\bibitem[{van Leeuwen(2007)}]{vanLeeuwen:2007dc}
van Leeuwen, F. 2007, A{\&}A, 474, 653

\bibitem[{Vigan {et~al.}(2016)Vigan, Bonnefoy, Ginski, Beust, Galicher, Janson,
  Baudino, Buenzli, Hagelberg, D'Orazi, Desidera, Maire, Gratton, Sauvage,
  Chauvin, Thalmann, Malo, Salter, Zurlo, Antichi, Baruffolo, Baudoz,
  Blanchard, Boccaletti, Beuzit, Carle, Claudi, Costille, Delboulb{\'e},
  Dohlen, Dominik, Feldt, Fusco, Gluck, Girard, Giro, Gry, Henning, Hubin,
  Hugot, Jaquet, Kasper, Lagrange, Langlois, Le~Mignant, Llored, Madec,
  Martinez, Mawet, Mesa, Milli, Mouillet, Moulin, Moutou, Orign{\'e}, Pavlov,
  Perret, Petit, Pragt, Puget, Rabou, Rochat, Roelfsema, Salasnich, Schmid,
  Sevin, Siebenmorgen, Smette, Stadler, Suarez, Turatto, Udry, Vakili, Wahhaj,
  Weber, \& Wildi}]{Vigan:2016gq}
Vigan, A., Bonnefoy, M., Ginski, C., {et~al.} 2016, A{\&}A, 587, A55

\bibitem[{Vogt {et~al.}(1994)Vogt, Allen, Bigelow, Bresee, Brown, Cantrall,
  Conrad, Couture, Delaney, Epps, Hilyard, Hilyard, Horn, Jern, Kanto, Keane,
  Kibrick, Lewis, Osborne, Pardeilhan, Pfister, Ricketts, Robinson, Stover,
  Tucker, Ward, \& Wei}]{Vogt:1994tb}
Vogt, S.~S., Allen, S.~L., Bigelow, B.~C., {et~al.} 1994, Proc. SPIE, 2198, 362

\bibitem[{Vogt {et~al.}(2014)Vogt, Radovan, Kibrick, Butler, Alcott, Allen,
  Arriagada, Bolte, Burt, Cabak, Chloros, Cowley, Deich, Dupraw, Earthman,
  Epps, Faber, Fischer, Gates, Hilyard, Holden, Johnston, Keiser, Kanto,
  Katsuki, Laiterman, Lanclos, Laughlin, Lewis, Lockwood, Lynam, Marcy, McLean,
  Miller, Misch, Peck, Pfister, Phillips, Rivera, Sandford, Saylor, Stover,
  Thompson, Walp, Ward, Wareham, Wei, \& Wright}]{Vogt:2014wxa}
Vogt, S.~S., Radovan, M., Kibrick, R., {et~al.} 2014, Publications of the
  Astronomical Society of the Pacific, 126, 359

\bibitem[{Wittenmyer {et~al.}(2011)Wittenmyer, Tinney, O'Toole, Jones, Butler,
  Carter, \& Bailey}]{Wittenmyer:2011fw}
Wittenmyer, R.~A., Tinney, C.~G., O'Toole, S.~J., {et~al.} 2011, The
  Astrophysical Journal, 727, 102

\bibitem[{Wittenmyer {et~al.}(2016)Wittenmyer, Butler, Tinney, Horner, Carter,
  Wright, Jones, Bailey, \& O'Toole}]{Wittenmyer:2016hp}
Wittenmyer, R.~A., Butler, R.~P., Tinney, C.~G., {et~al.} 2016, ApJ, 819, 1

\bibitem[{Wizinowich(2013)}]{Wizinowich:2013dz}
Wizinowich, P. 2013, Publications of the Astronomical Society of the Pacific,
  125, 798

\bibitem[{Yelda {et~al.}(2010)Yelda, Lu, Ghez, Clarkson, Anderson, Do, \&
  Matthews}]{Yelda:2010ig}
Yelda, S., Lu, J.~R., Ghez, A.~M., {et~al.} 2010, The Astrophysical Journal,
  725, 331

\bibitem[{Youdin(2011)}]{Youdin:2011gz}
Youdin, A.~N. 2011, ApJ, 742, 38

\bibitem[{Zechmeister {et~al.}(2013)Zechmeister, K{\"u}rster, Endl, Lo~Curto,
  Hartman, Nilsson, Henning, Hatzes, \& Cochran}]{Zechmeister:2013gt}
Zechmeister, M., K{\"u}rster, M., Endl, M., {et~al.} 2013, A{\&}A, 552, A78

\end{thebibliography}

\clearpage

\newpage

\end{document}